\documentstyle[prd,aps,epsfig]{revtex}

\begin{document}
%
%
\def\ov{\over}
\def\l{\left}
\def\r{\right}
\def\be{\begin{equation}}
\def\ee{\end{equation}}
\def\mib#1{{\bf #1}}
\def\ulap{\underline\Delta}
\def\unab{{\overline\nabla}}
\def\spose#1{\hbox to 0pt{#1\hss}}
\def\lta{\mathrel{\spose{\lower 3pt\hbox{$\mathchar"218$}}
     \raise 2.0pt\hbox{$\mathchar"13C$}}}
\def\gta{\mathrel{\spose{\lower 3pt\hbox{$\mathchar"218$}}
     \raise 2.0pt\hbox{$\mathchar"13E$}}}
\draft
\title{Quasiequilibrium sequences of synchronized and
       irrotational binary neutron stars
       in general relativity. I. Method and tests} 
\author{Eric Gourgoulhon, Philippe Grandcl\'ement, Keisuke Taniguchi,
        Jean-Alain Marck and Silvano Bonazzola}
\address{D\'epartement d'Astrophysique Relativiste et de Cosmologie \\
  UMR 8629 du C.N.R.S., Observatoire de Paris, \\
  F-92195 Meudon Cedex, France 
}
\date{15 December 2000}
\maketitle

\begin{abstract}

We present a numerical method to compute quasiequilibrium configurations
of close binary neutron stars in the pre-coalescing stage. 
A hydrodynamical treatment is performed under 
the assumption that the flow is either rigidly rotating or irrotational. 
The latter state is technically more 
complicated to treat than the former one (synchronized binary),
but is expected to represent fairly well the late evolutionary stages 
of a binary neutron star system. As regards the gravitational
field, an approximation of general relativity is used,
which amounts to solving five of the 
ten Einstein equations (conformally flat spatial metric).
The obtained system of partial differential
equations is solved by means of a multi-domain spectral method. 
Two spherical coordinate systems are introduced, one centered on
each star; this results in a precise description of the stellar
interiors. Thanks to the multi-domain approach, this high precision 
is extended to the strong field regions. The computational domain covers
the whole space so that exact boundary conditions are set to
infinity. Extensive tests of the numerical code are performed,
including comparisons with recent analytical solutions. Finally  
a constant baryon number sequence (evolutionary sequence)
is presented in details for a polytropic equation of state with $\gamma=2$. 
\end{abstract}



\section{Introduction} \label{s:intro}

Inspiraling neutron star binaries are expected to be among the
strongest sources of gravitational radiation that could be detected by
the interferometric detectors currently under construction (GEO600,
LIGO, and VIRGO) or in operation (TAMA300).  
Such binary systems are therefore subject to
numerous theoretical studies (see e.g. \cite{RasioS99} for a review).  
Among them there are 
(i) Post-Newtonian (PN) analytical treatments (e.g. \cite{BlancI98},
\cite{BuonaD99}, \cite{Tanig99}) and (ii)
fully relativistic hydrodynamical treatments, 
pioneered by the works of Oohara \&
Nakamura (see e.g.  \cite{OoharN97}), Wilson et al.
\cite{WilsoM95,WilsoMM96} and recently developed by 
Shibata \cite{Shiba99a,Shiba99b,Shiba99c,ShibaU00}, the Neutron Star
Grand Challenge group \cite{Suen99,FontMST00} 
and Oohara \& Nakamura \cite{OoharN99}.
These last three groups 
integrate forward in time the evolution equations
resulting from the 3+1 formulation of general relativity
\cite{York79,SeideS99}. 
In parallel of these dynamical calculations, some quasiequilibrium
formulation of the problem has been developed 
\cite{BonazGM97b,Asada98,Teuko98,Shiba98} and successfully implemented
\cite{BonazGM99a,MarroMW99,UryuE00,UryuSE00}. The basic assumption underlying
the quasiequilibrium calculations is that the timescale of the
orbit shrinking is larger than that of the orbital revolution
in the pre-coalescing state. 
Consequently the evolution of the binary system can be approximated by
a succession of exactly circular orbits, hence the name 
{\em quasiequilibrium}. The study of these quasiequilibrium configurations
is justified in the view that the fully dynamical computations mentioned
above are only in their infancy. In particular, they cannot follow
more than a few orbits. Also they involve a rather coarse resolution of the
stars, being performed in a single box with Cartesian coordinates. 
Another motivation for computing quasiequilibrium configurations 
is to provide valuable initial conditions for the dynamical evolutions
\cite{ShibaU00,Suen99,OoharN99}.

The first quasiequilibrium configurations of binary neutron stars in
general relativity have been obtained three years ago by  
Baumgarte et al.  \cite{BaumgCSST97,BaumgCSST98}, followed by
Marronetti et al. \cite{MarroMW98}. However these computations 
considered synchronized binaries. This rotation state 
does not correspond to physical situations, since
it has been shown that the gravitational-radiation driven evolution is
too rapid for the viscous forces to synchronize the spin of each
neutron star with the orbit \cite{Kocha92,BildsC92} as they do for
ordinary stellar binaries.  Rather, the viscosity is negligible and 
the fluid velocity circulation (with respect to some inertial
frame) is conserved in these systems.  Provided that the initial spins
are not in the millisecond regime, this means that close binary
configurations are well approximated by zero vorticity (i.e. {\em
irrotational}) states. Irrotational configurations are more
complicated to obtain because the fluid velocity does not vanish in 
the co-orbiting frame (as it does for synchronized binaries). 
We have successfully developed a numerical method to tackle this problem
and presented the first quasiequilibrium configurations 
of irrotational binary neutron stars elsewhere \cite{BonazGM99a}. 
The numerical technique relies on a multi-domain spectral method
\cite{BonazGM98a} within spherical coordinates. 
Since then, two other groups
have obtained relativistic irrotational configurations: 
(i) Marronetti, Mathews \& Wilson \cite{MarroMW99,MarroMW00} 
by means of single-domain finite difference
method within Cartesian coordinates and (ii) Uryu, Eriguchi 
and Shibata \cite{UryuE00,UryuSE00}
by means of multi-domain finite difference
method within spherical coordinates. 

The present article is devoted to the detailed presentation of our method,
along with numerous tests of the numerical code, while
the previous letter \cite{BonazGM99a} gave only a sketch of the equations
and some results about an evolutionary sequence built on a polytropic
equation of state. In particular, that letter focuses on the evolution 
of the central density along the sequence in order to investigate 
the stability of each star against gravitational collapse. That 
study was motivated by the 1995 finding of Wilson et al.
\cite{WilsoM95,WilsoMM96} that the neutron stars may
individually collapse into a black hole prior to merger. This
unexpected result has been called into question by a number of authors
(see Ref.~\cite{MatheMW98} for a summary of all the criticisms and
some answers). Recently Flanagan \cite{Flana99} has found an error in
the analytical formulation used by Wilson et al. \cite{WilsoM95,WilsoMM96}.
New numerical computations by Mathews and Wilson \cite{MatheW99},
using a corrected code, show a significantly reduced compression effect.  

The plan of the article is as follows. We start by presenting the
equations governing binary stars in general relativity in Sect.~\ref{s:rel_eq}
(hydrodynamics) and Sect.~\ref{s:grav} (gravitational field). 
The numerical method developed to integrate these equations is presented 
in Sect.~\ref{s:num}. Section~\ref{s:tests} is then devoted to the
tests passed by the numerical code. Astrophysical 
results are then presented 
in Sect.~\ref{s:results} for
an evolutionary sequence of irrotational binary stars constructed
on polytropic equation of state of adiabatic index $\gamma=2$.
Section~\ref{s:discus} contains the final discussion (comparison of our
method with that used by other groups, conclusions about the tests) and
future prospects. 

Throughout the present article, we use units of $G=c=1$ where
$G$ and $c$ denote the gravitational constant and speed of light.

\section{Relativistic equations governing binaries in circular orbits}
\label{s:rel_eq}

Our treatment of binary neutron stars relies on the assumptions
of (i) quasiequilibrium state (i.e. 
steady state in the co-orbiting frame), (ii) a specific 
velocity state for the fluid: either rigid or irrotational flow,
(iii) the spatial 3-metric is almost conformally flat.  
In this section, we examine the assumptions (i) and (ii), without 
invoking assumption (iii), which will be introduced only in 
Sect.~\ref{s:grav}.

\subsection{Quasiequilibrium assumption} \label{s:quasieq}

In the late inspiral phase, before any orbital instability or 
merging of the two stars, the evolution of binary neutron stars
can be approximated by a succession of circular orbits. Indeed
when the separation between the centers of the two neutron stars
is about $50{\rm\, km}$ (in harmonic coordinates)
the time variation of the orbital period, $\dot P_{\rm orb}$, computed at 
the 2nd Post-Newtonian (PN) order by means of the formulas established by 
Blanchet et al. \cite{BlancDIWW95} is about $2\%$.
The evolution at this stage can thus be still considered as a sequence
of equilibrium configurations.
Moreover the orbits are expected to be circular (vanishing eccentricity), 
as a consequence of the gravitational radiation reaction \cite{Peter64}.
In terms of the spacetime geometry, 
we translate these assumptions by
demanding that there exists a Killing vector field $\mib{l}$ which is
expressible as \cite{BonazGM97b}
\be \label{e:helicoidal}
 \mib{l} = \mib{k} + \Omega \, \mib{m} \ ,
\ee
where $\Omega$ is a constant, to be identified with the orbital
angular velocity with respect to a distant inertial observer, and
$\mib{k}$ and $\mib{m}$ are two vector fields with the following 
properties: $\mib{k}$ is timelike at least far from the binary and 
is normalized so that far from the star it coincides
with the 4-velocity of inertial observers with respect to which 
the total ADM 3-momentum of the system vanishes. 
On the other hand $\mib{m}$ is a spacelike vector field which
has closed orbits and is zero on a two dimensional timelike surface, called
the {\em rotation axis}. $\mib{m}$ is normalized so that 
$\nabla(\mib{m}\cdot\mib{m}) \cdot \nabla(\mib{m}\cdot\mib{m}) / 
(4 \, \mib{m}\cdot\mib{m})$
tends to $1$ on the rotation axis [this latter condition ensures that
the parameter $\varphi$ associated with $\mib{m}$ along its trajectories
by $\mib{m} = \partial/\partial \varphi$ has the standard $2\pi$
periodicity].
Let us call $\mib{l}$ the {\em helicoidal Killing vector}. We assume
that $\mib{l}$ is a symmetry generator not only for the spacetime
metric $\mib{g}$ but also for all the matter fields. In particular,
$\mib{l}$ is tangent to the world tubes representing the surface of 
each star, hence its qualification of {\em helicoidal} 
(cf. Figure~1 of Ref.~\cite{BonazGM97b}).

The approximation suggested above amounts to neglecting outgoing gravitational
radiation. For non-axisymmetric systems --- as binaries are ---
imposing $\mib{l}$ as an exact 
Killing vector leads to a spacetime which is not asymptotically flat
\cite{GibboS83}. Thus, in solving for the gravitational field equations,
a certain approximation has to be devised in order to 
avoid the divergence of some metric coefficients at infinity. For instance
such an approximation could be the Wilson \& Mathews scheme \cite{WilsoM89}
that amounts to solving only for the Hamiltonian and momentum constraint 
equations, as well as the trace of the spatial part of
the ``dynamical'' Einstein equations (cf. Sect.~\ref{s:conf_flat}). 
This approximation has been used in all the 
relativistic quasiequilibrium studies to date 
and is consistent 
with the existence of the helicoidal Killing vector field (\ref{e:helicoidal}).
Note also that since the gravitational radiation reaction shows up
only at the 2.5-PN order, the helicoidal symmetry is exact up to the
2-PN order.

Following the standard 3+1 formalism \cite{York79}, 
we introduce a foliation of spacetime  
by a family of spacelike hypersurfaces $\Sigma_t$ such that at spatial 
infinity, the vector $\mib{k}$ introduced in Eq.~(\ref{e:helicoidal})
is normal to $\Sigma_t$ and the ADM 3-momentum in $\Sigma_t$ vanishes 
(i.e. the time $t$ is the proper time of an asymptotic inertial observer
at rest with respect to the binary system). Asymptotically, 
$\mib{k} = \partial/\partial t$ and $\mib{m} = \partial/\partial\varphi$,
where $\varphi$ is the azimuthal coordinate associated with the above
asymptotic inertial observer, so that Eq.~(\ref{e:helicoidal}) can be
re-written as
\be \label{e:helicoidal_partial}
 \mib{l} = {\partial\over\partial t} 
		+ \Omega \, {\partial\over\partial \varphi} \ .
\ee

One can then introduce the 
shift vector $\mib{B}$ of co-orbiting coordinates by means of the 
orthogonal decomposition of $\mib{l}$ with respect to the $\Sigma_t$
foliation:
\be \label{e:helicoidal_n}
	\mib{l} = N \, \mib{n} - \mib{B} ,
\ee
where  
$\mib{n}$ is the unit future directed vector normal to $\Sigma_t$, 
$N$ is called the lapse function and $\mib{n}\cdot\mib{B}=0$.

\subsection{Fluid motion}

We consider a perfect fluid, which constitutes an excellent approximation 
for neutron star matter. The matter stress-energy tensor is then
\be \label{e:stress-energy}
   \mib{T} = (e+p) \, \mib{u} \otimes \mib{u} + p \, \mib{g} \ , 
\ee
$e$ being the fluid proper energy density, $p$ the fluid pressure,
$\mib{u}$ the fluid 4-velocity, and $\mib{g}$ the spacetime metric. 
A zero-temperature equation of state (EOS) is a very good approximation for 
neutron star matter. For such an EOS, the first law of 
thermodynamics gives rise to the following identity (Gibbs-Duhem relation):
\be \label{e:Gibbs-Duhem}
	{\nabla p \over e+p} = {1\over h} \nabla h \ ,
\ee
where $h$ is the fluid specific enthalpy:
\be
	h := {e+p \over m_{\rm B} n} ,
\ee
$n$ being the fluid baryon number density and $m_{\rm B}$ the mean baryon 
mass: $m_{\rm B} = 1.66\times 10^{-27} {\rm\ kg}$. 
Note that for our zero-temperature EOS, $m_{\rm B} \, h$ is equal to the
fluid chemical potential.

By means of the identity (\ref{e:Gibbs-Duhem}), it is straightforward to 
show that the classical momentum-energy conservation equation 
$\nabla\cdot \mib{T} = 0$ is equivalent to the set of two 
equations\cite{Lichn67,Carte79}:
\be \label{e:canon}
	\mib{u} \cdot (\nabla \wedge \mib{w}) = 0 \ ,
\ee
\be \label{e:baryon_conserv}
	\nabla\cdot (n \mib{u}) = 0 \ . 
\ee
In Eq.~(\ref{e:canon}), $\mib{w}$ is the co-momentum 1-form
\be \label{e:co-mom}
	\mib{w} := h \, \mib{u}
\ee
and $\nabla \wedge \mib{w}$ denotes the exterior derivative of $\mib{w}$, 
i.e. the vorticity 2-form\cite{Lichn67}. In terms of components, one has
\be
	(\nabla \wedge \mib{w})_{\alpha\beta} = \nabla_\alpha w_\beta
	- \nabla_\beta w_\alpha  = \partial_\alpha w_\beta
		- \partial_\beta w_\alpha \ . 
\ee

The vorticity 2-form enters Cartan's identity which states that the
Lie derivative of the 1-form $\mib{w}$ along the vector field $\mib{l}$
is 
\be \label{e:Cartan}
	\pounds_{\mib{l}} \mib{w} = \mib{l}\cdot(\nabla\wedge\mib{w})
	+ \nabla(\mib{l}\cdot\mib{w}) \ .  
\ee
Because of the assumed helicoidal symmetry, 
${\pounds_{\mib{l}}} \mib{w} = 0$,
so that Cartan's identity reduces to
\be \label{e:Cartan2}
	\mib{l}\cdot(\nabla\wedge\mib{w})
	+ \nabla(\mib{l}\cdot\mib{w}) = 0 \ .  
\ee
This equation reveals to be very useful in the following; this justifies
the introduction of the vorticity 2-form. 

In particular, performing the scalar product of Eq.~(\ref{e:Cartan2})
by the fluid 4-velocity $\mib{u}$ leads to 
\be
   \mib{l}\cdot(\nabla\wedge\mib{w})\cdot \mib{u}
	+ \mib{u}\cdot\nabla(\mib{l}\cdot\mib{w}) = 0 \ . 
\ee
The first term in the left-hand side vanishes by virtue of the equation 
of motion (\ref{e:canon}), so that we obtain
\be
     \mib{u}\cdot\nabla(\mib{l}\cdot\mib{w}) = 0 \ , 
\ee
which means that the quantity $\mib{l}\cdot\mib{w} = h \, \mib{l}\cdot \mib{u}$
is constant along each streamline. This is the
relativistic generalization of the classical Bernoulli theorem. 
At this stage, it must be noticed that, in order for the
constant to be uniform over the streamlines, i.e., to 
be a constant over spacetime, so that one gets
a first integral for the fluid motion,
some additional property of the flow must be required. 
In the following two sections, we explore two such additional properties:
rigidity and irrotationality.

\subsection{Rigid rotation} \label{s:rigid}

A rigid motion corresponds to {\em synchronized} stars (also called 
{\em corotating} stars).  
It is defined in relativity by the vanishing of the
expansion tensor 
$\theta_{\alpha\beta} := (g_\alpha^{\ \mu}+u_\alpha u^\mu) 
	(g_\beta^{\ \nu}+u_\beta u^\nu)
	\nabla_{(\nu} u_{\mu)}$ of the 4-velocity $\mib{u}$. In the
presence of a Killing vector $\mib{l}$, this can be realized by requiring
the colinearity of $\mib{u}$ and $\mib{l}$ : 
\be \label{e:rigid}
	\mib{u} = \lambda \, \mib{l} \ , 
\ee
where $\lambda$ is a scalar field related to the norm of $\mib{l}$ by
the normalization of the 4-velocity 
$\lambda = (-\mib{l}\cdot\mib{l})^{-1/2}$. 
Inserting relation (\ref{e:rigid}) into the equation of fluid motion 
(\ref{e:canon})
shows that the first term in Eq.~(\ref{e:Cartan2}) vanishes identically,
so that one gets the well known first integral of motion \cite{Boyer65}
\be \label{e:int_prem_rigid}
	\mib{l} \cdot \mib{w} = {\rm const.}
\ee

The second part of the equations of fluid motion, Eq.~(\ref{e:baryon_conserv})
(baryon number conservation), is trivially satisfied by the form
(\ref{e:rigid}) because $\mib{l}$ is a Killing vector.

\subsection{Irrotational flow} \label{s:irrot}

As recalled in Sect.~\ref{s:intro}, realistic binary neutron stars
are not expected to be in synchronized rotation, but rather to 
have an irrotational motion. A relativistic 
irrotational flow is defined by the vanishing of the vorticity
2-form \cite{Lichn67} :
\be \label{e:irrot}
	\nabla \wedge \mib{w} = 0 \ . 
\ee
This is equivalent to the existence of a scalar field $\Psi$ such that
\be \label{e:potential_flow}
	\mib{w} = \nabla \Psi \ .  
\ee
This is the relativistic definition of a {\em potential flow} \cite{LandaL89}. 
Note that the advantage of writing the equation for the fluid
motion in the form (\ref{e:canon})-(\ref{e:baryon_conserv})
rather than in the traditional form $\nabla\cdot \mib{T} = 0$ is that
one can see immediately that a flow of the form (\ref{e:potential_flow})
is a solution of (\ref{e:canon}). 

The second part of the equation of motion, Eq.~(\ref{e:baryon_conserv}),
is satisfied by the potential flow (\ref{e:potential_flow}) provided 
that $\Psi$ obeys to the equation
\be \label{e:eq_Psi_4d}
{n \over h} \nabla \cdot \nabla \Psi 
        + \nabla \Psi \cdot \nabla \left( {n \over h} \right) =0 \ .
\ee

Inserting the irrotationality condition (\ref{e:irrot}) into 
Eq.~(\ref{e:Cartan2}) results in an
equation showing the constancy of the scalar product $\mib{l}\cdot\mib{w}$:
\be \label{e:int_prem_irrot}
	\mib{l} \cdot \mib{w} = {\rm const.}
\ee
We therefore obtain the same first integral as in the rigid case 
(Eq.~(\ref{e:int_prem_rigid}) above). However note that the way to
get it is different: no use of the equation of motion (\ref{e:canon})
has been made to obtain (\ref{e:int_prem_irrot}), contrary to the
derivation of Eq.~(\ref{e:int_prem_rigid}).
The first integral (\ref{e:int_prem_rigid}) 
for rigid motion has been known for a long time,
at least since Boyer's work \cite{Boyer65}. To our knowledge, the version 
(\ref{e:int_prem_irrot}) for an irrotational flow in presence of a 
Killing vector is due to Carter \cite{Carte79}. 

\subsection{3+1 decomposition}

The first integral (\ref{e:int_prem_rigid}),(\ref{e:int_prem_irrot}),
common to both the rigid and irrotational motion, is expressed in
terms of the contraction of a spacetime
vector ($\mib{l}$) with a spacetime 1-form ($\mib{w}$). 
Going back to the 3+1 formalism mentioned in Sect.~\ref{s:quasieq},
let us re-express it in terms of quantities relative to the
hypersurfaces $\Sigma_t$. Following Ref.~\cite{BonazGM97b}, we
introduce the {\em co-orbiting observer}, whose 4-velocity $\mib{v}$ 
is the normalized symmetry generator: 
\be \label{e:v-co-orb}
	\mib{v} = (N^2 - \mib{B}\cdot\mib{B})^{-1/2}\,  \mib{l} \ ,
\ee
where the normalization factor has been deduced from 
Eq.~(\ref{e:helicoidal_n}). 
Note that in the rigid motion case, the co-orbiting observer
and the fluid comoving observer coincide: $\mib{u} = \mib{v}$
[cf. Eq.~(\ref{e:rigid})]. 
The 3+1 split of the 4-velocity $\mib{v}$ with respect to the Eulerian
observer is
\be \label{e:v-co-orb-2}
	\mib{v} = \Gamma_0 (\mib{n} + \mib{U}_0) \ ,
\ee
where 
\be \label{e:gamma_0}
	\Gamma_0 = - \mib{n} \cdot \mib{v} 
		 = (1-\mib{U}_0 \cdot \mib{U}_0)^{-1/2}
\ee 
is the Lorentz factor between
the two observers and $\mib{U}_0$ is the orbital 3-velocity with
respect to the Eulerian observer ($\mib{n}\cdot\mib{U}_0 = 0$). 
According to Eqs.~(\ref{e:v-co-orb}) 
and (\ref{e:helicoidal_n}), $\mib{U}_0$ is linked to the shift vector of
co-orbiting coordinates by
\be \label{e:U_0}
   \mib{U}_0 = - {\mib{B} \ov N} \ . 	
\ee
Thanks to the second part of Eq.~(\ref{e:gamma_0}), 
Eq.~(\ref{e:v-co-orb}) can be re-written as 
\be \label{e:v-co-orb-3}
  \mib{v} = {\Gamma_0 \over N} \, \mib{l} \ . 
\ee

The fluid motion can be described by
the following orthogonal decompositions of $\mib{u}$:
\be \label{e:u-decomp}
	\mib{u} = \Gamma (\mib{v} + \mib{V}) 
		= \Gamma_{\rm n} (\mib{n} + \mib{U})  \ , 
\ee
where $\Gamma = - \mib{v} \cdot \mib{u}$ 
(resp. $\Gamma_{\rm n} = - \mib{n} \cdot \mib{u}$)
is the Lorentz factor between the fluid and
the co-orbiting (resp. Eulerian) observer, and 
$\mib{V}$ (resp. $\mib{U}$) is the fluid 3-velocity with 
respect to the co-orbiting (resp. Eulerian) observer. 
In particular, $\mib{v}\cdot\mib{V} = 0$, $\mib{n}\cdot\mib{U} = 0$
and
\be \label{e:def_U}
	\mib{U} = {1\ov \Gamma_{\rm n}} \, \mib{h} \cdot \mib{u} \ ,
\ee
where 
\be
	\mib{h} := \mib{g} + \mib{n} \otimes \mib{n}  
\ee
is the orthogonal projector onto the spatial hypersurfaces $\Sigma_t$;
$\mib{h}$ can also be viewed as the metric induced by $\mib{g}$ onto
the hypersurfaces $\Sigma_t$. 
Performing the scalar product of Eq.~(\ref{e:v-co-orb-2}) with the
second part of Eq.~(\ref{e:u-decomp}) leads to an expression of the Lorentz
factor $\Gamma$ in terms of quantities relative to the Eulerian
observer only:
\be \label{e:gamma_3p1}
	\Gamma = \Gamma_{\rm n} \Gamma_0 ( 1 - \mib{U} \cdot \mib{U}_0 ) \ . 
\ee
Similarly, performing the projection of the second part of 
Eq.~(\ref{e:u-decomp}) onto the hyperplane orthogonal to $\mib{v}$
results in the expression of 
the fluid 3-velocity $\mib{V}$ with respect to the co-orbiting
observer in terms of the 3-velocities $\mib{U}$ and
$\mib{U}_0$, both defined with respect to the Eulerian observer:
\be \label{e:3-vitV}
	\mib{V} = {\Gamma_0\over 1 - \mib{U}\cdot \mib{U}_0}
	\left[ \mib{U}_0 \cdot ( \mib{U} - \mib{U}_0 ) \mib{n}
		+ \mib{U} - \mib{U}_0 
		+ (\mib{U}\cdot\mib{U}_0)\, \mib{U}_0
		- (\mib{U}_0\cdot\mib{U}_0)\, \mib{U} \right] \ . 
\ee
Note that in the case where $\mib{U}$ and $\mib{U}_0$ are aligned 
($\mib{U} = U \mib{e}$ and $\mib{U}_0 = U_0 \mib{e}$, $\mib{e}$ being
some unit vector in $\Sigma_t$) relation (\ref{e:3-vitV}) reduces to 
the classical velocity-addition law of special relativity:
$\mib{V} = (U-U_0)/(1-U U_0) \, \mib{e'}$, where 
$\mib{e'} = \Gamma_0(\mib{e} + U_0 \mib{n})$ is the unit vector 
deduced from $\mib{e}$ by a boost of velocity $U_0$.
In particular for $\mib{U} = \mib{U}_0$, which corresponds to
synchronized binaries, $\mib{V}$ vanishes identically. 

For irrotational binaries, $\mib{U}$ 
is related to the potential $\Psi$ by combining
Eqs.~(\ref{e:co-mom}), (\ref{e:potential_flow}) and (\ref{e:def_U}) :
\be \label{e:U=grad_psi}
	\mib{U} = {1\over \Gamma_{\rm n} h} \, \mib{D} \Psi \ ,
\ee
where $\mib{D}$ is the covariant derivative associated with the metric
$\mib{h}$ of spatial hypersurfaces $\Sigma_t$. Combined with the relation
$\Gamma_{\rm n} = (1-\mib{U}\cdot\mib{U})^{-1/2}$, this relation results
in
\be \label{e:Gamma_n_cov}
	\Gamma_{\rm n} = \left( 1 + {1 \over h^2} \, \mib{D} \Psi 
			\cdot \mib{D} \Psi \right) ^{1/2} \ . 
\ee

We are now in position to write the 3+1 form of the
first integral (\ref{e:int_prem_rigid}),(\ref{e:int_prem_irrot}),
common to both the rigid and irrotational motion. 
Substituting relation (\ref{e:co-mom}) for $\mib{w}$ and relation
(\ref{e:v-co-orb-3}) for $\mib{l}$ into Eq.~(\ref{e:int_prem_rigid})
results in
\be \label{e:int_prem_3p1_non_log}
	h N \, {\Gamma\over\Gamma_0} = {\rm const} \ . 
\ee
We shall use actually the logarithm of this relation:
\be \label{e:int_prem_3p1}
	H + \nu - \ln \Gamma_0 + \ln \Gamma = {\rm const} \ ,
\ee
with the following definitions:
\be
	H := \ln h 
\ee
and
\be \label{e:def_nu}
	\nu := \ln N \ . 
\ee
These two quantities have immediate meaning at the Newtonian limit: $H$ is
the (non-relativistic) specific enthalpy and $\nu$ is the Newtonian 
gravitational potential.
The first integral of motion written as (\ref{e:int_prem_3p1}) coincides
with Eq.~(66) of Ref.~\cite{BonazGM97b}. The link with the alternative
expressions
derived by Teukolsky \cite{Teuko98} and by Shibata \cite{Shiba98} for
the irrotational case is performed in Appendix \ref{s:app_link}. 
Note that $\ln\Gamma = 0$ for synchronized binaries, so that 
Eq.~(\ref{e:int_prem_3p1}) simplifies somewhat. 
Note also that substituting Eq.~(\ref{e:gamma_3p1}) for $\Gamma$ in
Eq.~(\ref{e:int_prem_3p1}) leads to an alternative expression of the
first integral of motion which contains only quantities relative
to the Eulerian observer:
\be \label{e:int_prem_3p1-var}
	H + \nu + \ln \Gamma_{\rm n} + \ln ( 1 - \mib{U} \cdot \mib{U}_0 ) 
							= {\rm const} \ .
\ee
However in the following, we shall use only the form (\ref{e:int_prem_3p1}).

Let us now turn to the 3+1 form the differential equation (\ref{e:eq_Psi_4d})
for the velocity potential $\Psi$ of irrotational flows. 
Taking into account the helicoidal 
symmetry, Eq.~(\ref{e:eq_Psi_4d}) becomes
\be \label{e:psicov}
n \, \mib{D}\cdot \mib{D} \Psi + \mib{D} n \cdot \mib{D} \Psi 
  = h \Gamma_{\rm n} \, \mib{U}_0 \cdot \mib{D} n
  + n \left(   \mib{D} \Psi \cdot  \mib{D} \ln {h\over N}
     + \mib{U}_0 \cdot \mib{D} \Gamma_{\rm n} \right)
  + n h K \Gamma_{\rm n} 
\ ,
\ee
where $K$ is the
trace of the extrinsic curvature tensor of the $\Sigma_t$ hypersurfaces.
This equation has been obtained by Teukolsky \cite{Teuko98} and
independently by Shibata \cite{Shiba98}. We refer to these authors for the
details of the derivation of Eq.~(\ref{e:psicov}) from Eq.~(\ref{e:eq_Psi_4d}).

\subsection{Newtonian limit}

At the Newtonian limit, the Eulerian observer is an inertial
observer. Eqs.~(\ref{e:helicoidal_partial}) and 
(\ref{e:helicoidal_n}) show that 
$\mib{B} = - \Omega {\partial\over\partial \varphi}$, so that
Eq.~(\ref{e:U_0}) for the velocity of the co-orbiting observer with respect
to the inertial observer becomes
\be
	\mib{U}_0 = \Omega \times \mib{r} \ , 
\ee
where $\mib{r}$ denotes the position vector with respect to the 
center of mass of the system. The logarithm of the corresponding Lorentz factor
tends to (minus) the centrifugal potential [cf. Eq.~(\ref{e:gamma_0})]
\be
	\ln \Gamma_0 = {1\over 2} (\Omega \times \mib{r})^2 \ . 
\ee
The Newtonian limit of the first integral of motion (\ref{e:int_prem_3p1})
for synchronized binaries ($\ln\Gamma = 0$) gives the classical
expression
\be
    H + \nu - {1\over 2} (\Omega \times \mib{r})^2 = {\rm const} \ ,
\ee
where, as recalled above, $H$ is the fluid specific enthalpy and $\nu$
the Newtonian gravitational potential. 

In the irrotational case, the Newtonian limit results in the following
fluid velocity with respect to the inertial frame [set $h=1$ and 
$\Gamma_{\rm n}=1$ in Eq.~(\ref{e:U=grad_psi})]
\be
	\mib{U} = \vec{\nabla} \Psi \ , 
\ee
where $\vec{\nabla}$ denotes the standard 3-dimensional gradient
operator, i.e.
the Newtonian limit of the operator $\mib{D}$ introduced above. 
The first integral of motion (\ref{e:int_prem_3p1-var}) reduces then to
\be
	H + \nu + {1\over 2} (\vec{\nabla} \Psi)^2
	- (\Omega \times \mib{r})\cdot \vec{\nabla} \Psi = {\rm const}.
\ee
We recognize the classical expression (compare e.g. with Eq.~(12) of 
Ref.~\cite{Teuko98} or Eq.~(11) of Ref.~\cite{UryuE98b}). 
The Newtonian limit of the continuity equation (\ref{e:psicov}) reads
\be 
   n \Delta \Psi + \vec{\nabla} n \cdot \vec{\nabla} \Psi =
	(\Omega\times\mib{r}) \cdot \vec{\nabla} n \ .   
\ee
Here again, we recognize the classical expression (compare e.g. with 
Eq.~(13) of Ref.~\cite{Teuko98}).

\section{Gravitational field equations} \label{s:grav}

\subsection{A simplifying assumption: the conformally flat 3-metric}
\label{s:conf_flat}

As a first step in the treatment of binary configurations in
general relativity, a simplifying assumption can be introduced, in
order to reduce the computational task, namely to take the 3-metric 
induced in the hypersurfaces $\Sigma_t$ to be conformally flat:
\be \label{e:conf_plat}
	\mib{h} = A^2 \, \mib{f} \ , 
\ee
where $A$ is some scalar field and $\mib{f}$ is a flat 3-metric. 
This assumption has been  first introduced by Wilson \& Mathews \cite{WilsoM89}
and has been employed in all the studies of 
quasiequilibrium relativistic binaries to date 
\cite{BonazGM99a,MarroMW99,UryuE00,UryuSE00,BaumgCSST97,BaumgCSST98}. 
It has been also used in binary black hole initial data computations
(see e.g. \cite{Cook94,BrandB97,Baumg00,PfeifTC00}). 
It is physically less justified than the assumption of quasiequilibrium 
discussed above. However, some possible justifications of 
(\ref{e:conf_plat}) are
\begin{enumerate}
\item it is exact for spherically symmetric configurations;
\item it is very accurate for axisymmetric rotating neutron 
stars \cite{CookST96};
\item the 1-PN metric fits it;
\item the 2.5-PN metric \cite{BlancFP98} deviates from it by only 2 \% 
for two $1.4 \, M_\odot$ 
neutron stars as close as
30~km (in harmonic coordinates) \cite{BonazGM00}.
\end{enumerate}

\subsection{Partial differential equations for the metric}

To benefit from the helicoidal symmetry, 
we use co-orbiting coordinates $(t,x^1,x^2,x^3)$, i.e. coordinates
adapted to the Killing vector $\mib{l}$: ${\partial/\partial t} = \mib{l}$. 
Assuming the conformally flat form (\ref{e:conf_plat})
for $\mib{h}$, the full spacetime metric takes then the 
form\footnote{Latin indices $i,j,\ldots$ run from 1 to 3.}
\begin{equation} \label{e:met}
ds^2 = -(N^2 - B_i B^i) dt^2 - 2 B_i dt\, dx^i 
        + A^2 f_{ij} dx^i\, dx^j \ .
\end{equation}
We have thus five metric functions to determine: the lapse $N$, the
conformal factor $A$ and the three components $B^i$ of the shift vector
$\mib{B}$ [see Eq.~(\ref{e:helicoidal_n})].
Let us define auxiliary metric quantities: we have already
introduced the logarithm of $N$, $\nu$, via
Eq.~(\ref{e:def_nu}); we introduce now the shift vector
of non-rotating coordinates:
\be \label{e:shift_non_rot}
	\mib{N} = \mib{B} + \Omega {\partial\over\partial \varphi} \ , 
\ee
and the quantity 
\be
	\beta := \ln(AN) . 
\ee
At the Newtonian limit $\mib{N}=0$ and $\beta=0$. 

In the following, we choose the slicing of spacetime by the
hypersurfaces $\Sigma_t$ to be maximal. This results in $K=0$. 

The Killing equation $\nabla_\alpha l_\beta + \nabla_\beta l_\alpha = 0$,
gives rise to a relation between the $\Sigma_t$ extrinsic curvature tensor
$\mib{K}$ and the shift vector $\mib{N}$ (via Eq.~(\ref{e:helicoidal_n})
and the relation $\nabla \mib{n} = - \mib{K} - \mib{n}\otimes \mib{D}\ln N$)
\be \label{e:kij}
K^{ij} = - {1 \over 2 N}
          \left(   D^i B^j + D^j B^i \right)
	=  - {1 \over 2 A^2 N}
          \left\{
                \unab^i N^j + \unab^j N^i
                  - {2 \over 3} f^{ij} \unab_k N^k 
          \right\} \ ,
\ee
where $\unab$ stands for the covariant derivative associated with the flat
3-metric $\mib{f}$. Here and in the following, the index $i$ of $\unab^i$
is supposed to be raised with the metric $\mib{f}$.
Note that since $\partial/\partial\phi$ is a Killing
vector of the flat metric $\mib{f}$, the second part of
this  equation stands also with 
$N^i$ replaced by $B^i$.

The trace of the spatial part of the Einstein equation, combined with the
Hamiltonian constraint equation, result in the following two equations
\be \label{e:nu}
\ulap \nu = 4 \pi A^2 (E + S) + A^2 K_{ij} K^{ij} - \unab_i \nu \unab^i \beta 
   \ ,
\ee
\be \label{e:beta}
\ulap \beta = 4 \pi A^2 S + {3 \over 4} A^2 K_{ij} K^{ij} 
        - {1 \over 2} \left( \unab_i \nu \unab^i \nu  
        		+ \unab_i \beta \unab^i \beta \right) \ ,
\ee
whereas the momentum constraint equation yields, by means of Eq.~(\ref{e:kij}),
\be \label{e:shift}
\ulap N^i + {1 \over 3} \unab^i \left( \unab_j N^j \right) =
        - 16 \pi N A^2 (E + p) U^i + 2 N A^2 K^{ij} \unab_j (3 \beta - 4 \nu)
	\ . 
\ee
In these equations, 
$\ulap := \unab^i \unab_i$ is the Laplacian operator associated with 
the flat metric $\mib{f}$, and $E$ and $S$ are respectively the matter energy 
density and trace of the stress tensor, both as measured 
by the Eulerian observer:
\be \label{e:ener_euler}
        E := \mib{n}\cdot\mib{T}\cdot\mib{n} = \Gamma_{\rm n}^2 (e+p) - p,
\ee
\be \label{e:s_euler}
	S := \mib{h}\cdot\mib{T} = 3p +  (E +p) \mib{U}\cdot \mib{U} \ .  
\ee

The equations to be solved to get the metric coefficients are the 
elliptic equations (\ref{e:nu})-(\ref{e:shift}). Note that they 
represent only five of the ten Einstein equations. The remaining five Einstein
equations are not used in this procedure. Moreover, some of
these remaining equations
are certainly violated, reflecting the fact that the conformally flat 
3-metric (\ref{e:conf_plat}) is an approximation to the exact metric
generated by a binary system. 

At the Newtonian limit, Eqs.~(\ref{e:beta}) and (\ref{e:shift}) reduce 
to $0=0$. There remains only Eq.~(\ref{e:nu}), which gives the usual
Poisson equation for the gravitational potential $\nu$. 

\subsection{Equations for the fluid with a conformally flat 3-metric}

In this section we explicitly write some equations for fluid quantities
when the 3-metric takes the form (\ref{e:conf_plat}).
First the Lorentz factor (\ref{e:gamma_0})
between the co-orbiting and Eulerian observers writes
\be
	\Gamma_0 = (1- A^2 f_{ij} U_0^i U_0^j)^{-1/2} \ . 
\ee

For irrotational motion, the expression (\ref{e:Gamma_n_cov}) for the
Lorentz factor $\Gamma_{\rm n}$ between the fluid and Eulerian observers 
becomes
\be \label{e:gamma_n_flat}
	\Gamma_{\rm n} = \left( 1 + {1 \over A^2 h^2} \, f^{ij}
		\unab_i \Psi \unab_j \Psi \right) ^{1/2} \ .  
\ee
The corresponding fluid 3-velocity (\ref{e:U=grad_psi}) is
\be \label{e:U_flat_grad_Psi}
	U^i = {1\over A^2 \Gamma_{\rm n} h} \unab^i \Psi \ . 
\ee 
The Lorentz factor $\Gamma$ between the fluid and co-orbiting observer,
which enters in the first integral of motion (\ref{e:int_prem_3p1}),
is deduced from the above quantities via Eq.~(\ref{e:gamma_3p1}) :
\be \label{e:gamma_flat}
	\Gamma = \Gamma_{\rm n} \Gamma_0 (1- A^2 f_{ij} U^i U_0^j) \ . 
\ee

Let us now consider the continuity equation (\ref{e:psicov}). 
For a zero-temperature EOS, $H$ can be considered as a function of 
the baryon density $n$ solely, so that one can introduce the thermodynamical
coefficient
\be
	\zeta := {d\ln H \over d\ln n} \ .  
\ee
The gradient of $n$ which appears in Eq.~(\ref{e:psicov}) can be then
replaced by a gradient of $H$ so that, using the metric (\ref{e:conf_plat}),
one obtains
\be \label{e:psinum}
\zeta H \ulap \Psi + \unab^i H \unab_i \Psi =
  A^2 h \Gamma_{\rm n} U_0^i \unab_i H
 + \zeta H \left(
        \unab^i \Psi \unab_i(H-\beta) 
        + A^2 h \, U_0^i \unab_i \Gamma_{\rm n}
                \right) \ .
\ee
The potential $\Psi$ is in fact dominated by a pure translational part. 
Therefore, we write, in each star,
\be \label{e:Psi_Psi0}
	\Psi =: \Psi_0 + f_{ij} W^i_0 x^j \ , 
\ee 
where $W^i_0$ is the constant (translational) velocity field defined
as the central\footnote{The centers of the stars are defined in 
Sect.~\ref{s:domains}.} value of 
\be
	W^i := A^2 h \Gamma_{\rm n} U_0^i \ . 
\ee
Then $\unab^i \Psi = \unab^i \Psi_0 + W^i_0$ and $\ulap \Psi = \ulap \Psi_0$,
so that Eq.~(\ref{e:psinum}) becomes
\be \label{e:psinum2}
\zeta H \ulap \Psi_0 + \left[ (1-\zeta H) \unab^i H + \zeta H \unab^i\beta
				 \right]
		\unab_i \Psi_0 =
     (W^i - W^i_0) \unab_i H
 + \zeta H \left( W^i_0 \unab_i(H-\beta) + 
  		{W^i\over \Gamma_{\rm n}} \unab_i \Gamma_{\rm n}
                \right) \ . 
\ee
The advantage to solve Eq.~(\ref{e:psinum2}) instead of Eq.~(\ref{e:psinum})
is that the right-hand side of the former is much smaller than the 
right-hand side of the latter, due to the factor $W^i - W^i_0$, instead
of $W^i$, in front of $\unab_i H$. 

\subsection{Global quantities} \label{s:global}

The total mass-energy content in a $\Sigma_t$ hypersurface is given by
the Arnowitt-Deser-Misner (ADM) mass $M$, which is expressed by means of the 
surface integral at spatial infinity
\be
    M = {1\over 16\pi} \oint_\infty f^{ik} f^{jl} \left(
	  \unab_j h_{kl} - \unab_k h_{jl} \right) \, dS_i  
\ee
(see e.g. Eq.~(20.9) of Ref.~\cite{MisneTW73}). 
In the case of the conformally flat 3-metric $h_{ij} = A^2 f_{ij}$, this
integral can be written
\be \label{e:M_ADM_1}
   M = - {1\over 2\pi} \oint_\infty \unab^i A^{1/2} \, dS_i \ . 
\ee
By means of the Gauss-Ostragradsky formula, this expression can be converted
into the volume integral of $\ulap A^{1/2}$. This last quantity can be
expressed by subtracting Eq.~(\ref{e:nu}) from Eq.~(\ref{e:beta}) (recall that
$A=\exp(\beta-\nu)$), so that Eq.~(\ref{e:M_ADM_1}) becomes an integral
containing the matter energy density and the extrinsic curvature of
$\Sigma_t$:
\be
  M = \int_{\Sigma_t} A^{5/2} \left( E + {1\over 16\pi} K_{ij} K^{ij} \right)
			\, d^3 x \ . 
\ee

Following Bowen \& York \cite{BowenY80}
we define the total angular momentum in a $\Sigma_t$ hypersurface 
as the surface integral at spatial infinity\footnote{Note that contrary
to the ADM mass, the total angular momentum hence defined is not 
asymptotically gauge invariant: it is defined merely as the 
$1/r^3$ part of $K_{ij}$ {\em within our coordinates};
see York \cite{York80} for a discussion.}

\be
  J_i = {1\over 16\pi} \epsilon_{ijk} \oint_\infty \left(
	x^j K^{kl} - x^k K^{jl} \right) \, dS_l 
      = {1\over 16\pi} \epsilon_{ijk} \oint_\infty \left(
	x^j A^5 K^{kl} - x^k A^5 K^{jl} \right) \, dS_l \ , 
\ee
where $\epsilon_{ijk}$ is the 3-dimensional alternating tensor,
$x^i$ are Cartesian coordinates, and the second equality follows from the
fact that $A=1$ at spatial infinity. As for $M$, this integral can be
converted to a volume integral on $\Sigma_t$. Using the momentum
constraint equation $D_l K^{kl} = 8\pi (E+p)U^k$ and the fact that
$D_l K^{kl} = A^{-5} \partial(A^5 K^{kl})/\partial x^l$ for the
conformally flat 3-metric (\ref{e:conf_plat}), one obtains the expression
\be
   J_i = \epsilon_{ijk} \int_{\Sigma_t} A^5 (E+p) \, x^j U^k \, d^3 x \ . 
\ee

The baryon mass of each star is given by the integral on $\Sigma_t$ 
of the baryon number
density as measured by the Eulerian observer: $- n \mib{u}\cdot \mib{n} =
\Gamma_{\rm n} n$. In the case of the conformally flat 3-metric 
(\ref{e:conf_plat}), this integral becomes
\be \label{e:m_bar}
    M_{\rm B}^{\langle a \rangle} = m_{\rm B} 
		\int_{{\rm star}\ a} A^3 \Gamma_{\rm n} n \, d^3 x \ ,
			a=1,2 \ .  
\ee

\section{Numerical method} \label{s:num}

The equations to be solved to get a relativistic binary system
in quasiequilibrium
are the elliptic equations (\ref{e:nu})-(\ref{e:shift}) for the gravitational 
field, supplemented by the 
elliptic equation (\ref{e:psinum2}) for the velocity potential $\Psi_0$
in the irrotational case. A cold matter equation of state, of the
form
\be \label{e:eos}
	n=n(H) \qquad e=e(H) \qquad p=p(H) \ , 
\ee 
must be supplied to close the system of equations. The thermodynamical
quantity $H$ has been privileged in the EOS setting (\ref{e:eos})
because it is that quantity which is involved in the first integral
of motion (\ref{e:int_prem_3p1}).
Altogether, these equations constitute a system of coupled non-linear
partial differential equations. We solve this system by means of an iterative
procedure.  

\begin{figure}
\centerline{ \epsfig{figure=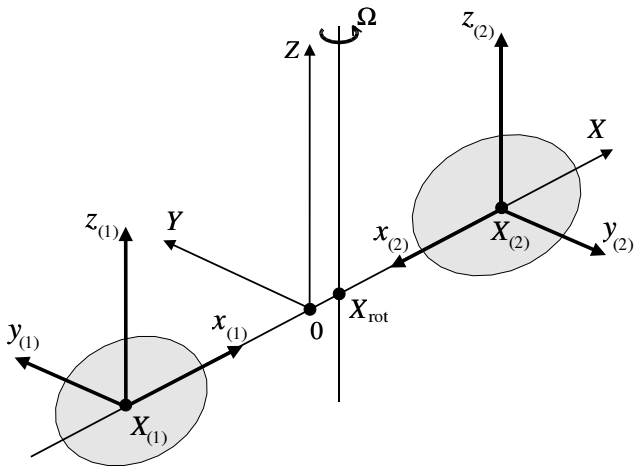,height=7cm} }
\caption[]{\label{f:coord} 
Coordinate systems used in the calculation.}  
\end{figure}

\subsection{Coordinate systems and computational domains} \label{s:domains}

We use co-orbiting coordinates
$(t,X,Y,Z)$ of Cartesian type  (i.e. $f_{ij} = \delta_{ij}$),
so that the line element (\ref{e:met}) can be written
\be
	ds^2 = - N^2 dt^2 + A^2 \left[ (dX-B^X dt)^2 + (dY-B^Y dt)^2 
						+ (dZ-B^Z dt)^2 \right] \ . 
\ee
In these coordinates, the two stars have fixed locations and figures.
Let us define the {\em center} of star no. $a$ ($a=1,2$) as the location of
the maximum enthalpy $H$ (or equivalently maximum density $e$) in
star $a$. Note that this {\em center} does not coincide with the 
center of mass of star $a$. 
We choose the coordinates $(X,Y,Z)$ such that (i) the orbital plane
is defined by $Z=0$, (ii) the two stellar centers are located along the
$X$ axis and (iii) the rotation axis (cf. Sect.~\ref{s:quasieq})
is located at $X=0$, $Y=0$. Let us then denote by $X_{\langle 1 \rangle}$ and $X_{\langle 2 \rangle}$ the
$X$ coordinates of the two stellar centers\footnote{In all this article,
indices or superscripts in $\langle\rangle$ will label the two stars.}.

In order to describe properly the stellar interiors, we introduce two
systems of Cartesian coordinates $(x_{\langle a \rangle},y_{\langle a
\rangle},z_{\langle a \rangle})$ centered on the two stars by (see
Fig.~\ref{f:coord}) \be \label{e:x-X}
	\left\{ \begin{array}{lcl}
		 x_{\langle 1 \rangle} & = & X - X_{\langle 1 \rangle}
		 \\
			 y_{\langle 1 \rangle} & = & Y \\ z_{\langle 1
			 \rangle} & = & Z \\
		\end{array} \right.  \qquad \mbox{and} \qquad \left\{
	\begin{array}{lcl}
		 x_{\langle 2 \rangle} & = & - (X - X_{\langle 2
		 \rangle}) \\
			 y_{\langle 2 \rangle} & = & -  Y \\ z_{\langle
			 2 \rangle} & = & Z  \\
		\end{array} \right.
\ee 
Note that the system $(x_{\langle 1 \rangle},y_{\langle 1
\rangle},z_{\langle 1 \rangle})$ is aligned with $(X,Y,Z)$, whereas
$(x_{\langle 2 \rangle},y_{\langle 2 \rangle},z_{\langle 2 \rangle})$
is anti-aligned (rotation of angle $\pi$ in the
 $(X,Y)$ plane) with $(X,Y,Z)$. This choice ensures that the companion
of star no. $a$ is located at $x_{\langle a \rangle} > 0$ for both
stars.  In particular, for equal mass stars, the descriptions of each
star in terms of $(x_{\langle a \rangle},y_{\langle a
\rangle},z_{\langle a \rangle})$ are identical. Furthermore we
introduce spherical coordinates $(r_{\langle a \rangle},\theta_{\langle
a \rangle},\varphi_{\langle a \rangle})$ ($a=1,2$) associated with each
of the Cartesian coordinate systems $(x_{\langle a \rangle},y_{\langle a
\rangle},z_{\langle a \rangle})$ by means of the usual formul\ae.

\begin{figure}
\centerline{ \epsfig{figure=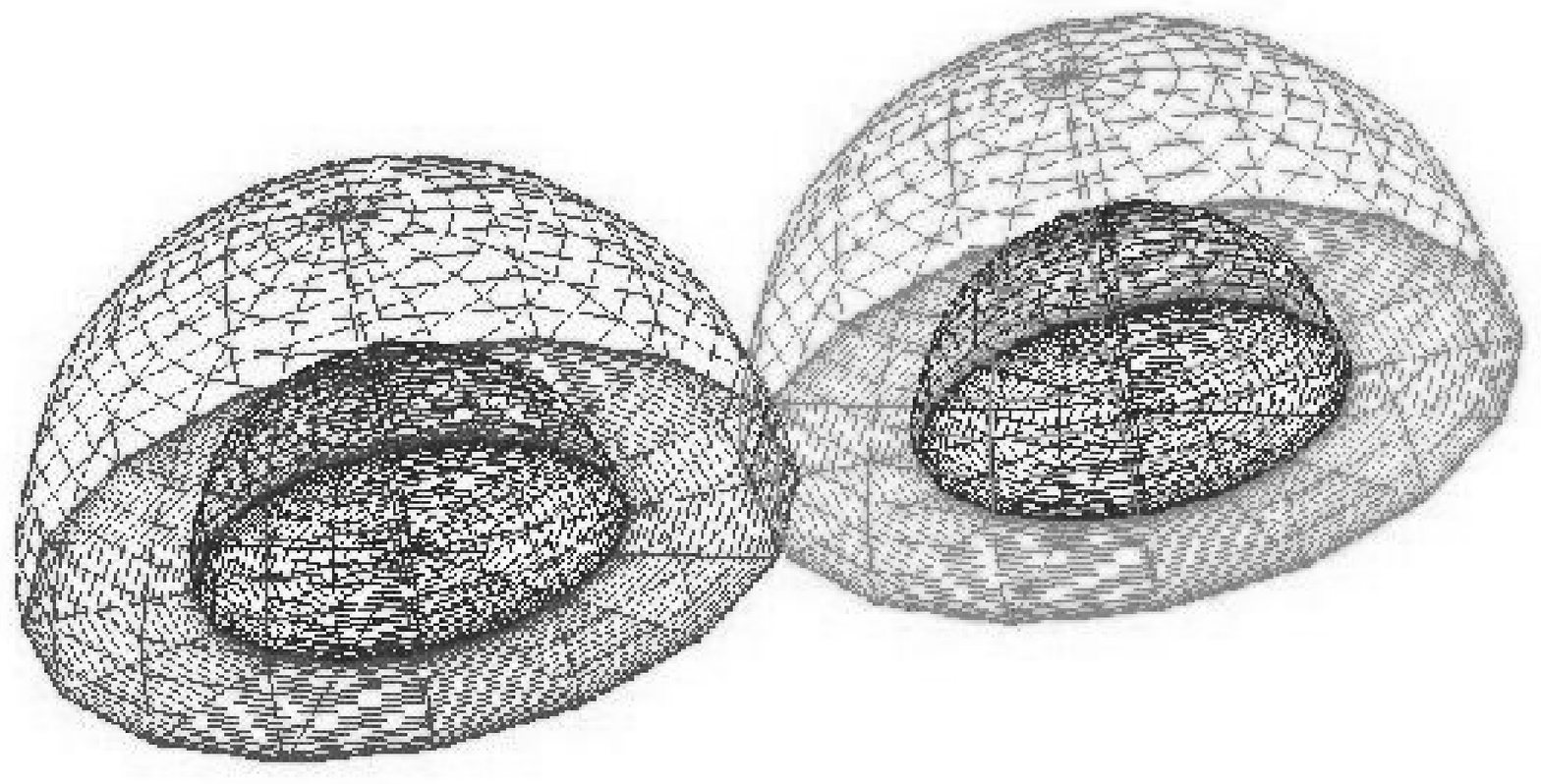,height=7cm} }
\caption[]{\label{f:domains} 
Domains used in the numerical computations, when $N_{\langle 1 \rangle} = N_{\langle 2 \rangle} = 3$.
The boundaries of domains ${\cal D}_0^{\langle 1 \rangle}$, ${\cal D}_1^{\langle 1 \rangle}$,
${\cal D}_0^{\langle 2 \rangle}$ and ${\cal D}_1^{\langle 2 \rangle}$ are represented. The outer 
boundaries of domains ${\cal D}_2^{\langle 1 \rangle}$ and ${\cal D}_2^{\langle 2 \rangle}$ are
located at infinity and are therefore not plotted.}  
\end{figure}

Since some of the equations to be solved are elliptic equations with
non-compactly supported sources, the computational domain must extend 
up to spatial infinity, i.e. must cover the full hypersurface $\Sigma_t$,
in order to put correct boundary conditions (flat spacetime). Any 
truncated computational domain (``box'') would result in approximate boundary
conditions, which inevitably would induce some error in the numerical
solution. The technique to cover the full $\Sigma_t$ is to divide it
in various domains, the outermost of it being compactified in order to
deal with finite computational domain only \cite{BonazGM98a}. Following the
introduction of the two coordinate systems 
$(r_{\langle a \rangle},\theta_{\langle a \rangle},\varphi_{\langle a \rangle})$
(one centered on each star), we will actually use two sets of such domains:
one centered on star no. 1, the other on star no. 2. The number 
$N_{\langle a \rangle}$ of domains in each set is arbitrary, being simply equal or larger
than 3. The list of the $N_{\langle 1 \rangle}+N_{\langle 2 \rangle}$ computational domains is 
\begin{eqnarray}
   & &	{\cal D}_0^{\langle 1 \rangle} \quad \ldots \quad {\cal D}_{M_{\langle 1 \rangle}-1}^{\langle 1 \rangle}
		\quad \ldots \quad {\cal D}_{N_{\langle 1 \rangle}-1}^{\langle 1 \rangle}  \nonumber \\
   & &	{\cal D}_0^{\langle 2 \rangle} \quad \ldots \quad  {\cal D}_{M_{\langle 2 \rangle}-1}^{\langle 2 \rangle}
		\quad \ldots \quad {\cal D}_{N_{\langle 2 \rangle}-1}^{\langle 2 \rangle}  \nonumber 
\end{eqnarray}
where 
\begin{itemize}
\item Domain ${\cal D}_0^{\langle a \rangle}$ ($a=1,2$) has the topology of a ball and
      contains the center of star $a$; it is designed thereafter as the
      {\em nucleus}.
\item $M_{\langle a \rangle}$ ($a=1,2$) is the number of domains which cover
 the interior of star $a$. It obeys 
$M_{\langle a \rangle} \geq 1$ and $M_{\langle a \rangle} \leq N_{\langle a \rangle}-2$.
The outer boundary of domain ${\cal D}_{M_{\langle a \rangle}-1}^{\langle a \rangle}$ coincides exactly
with the surface of star $a$. The topology of domains 
${\cal D}_1^{\langle a \rangle},\ldots,{\cal D}_{M_{\langle a \rangle}-1}^{\langle a \rangle}$ is that of a spherical
shell; these domains are designed thereafter as the {\em shells}. 
\item Domains ${\cal D}_{M_{\langle a \rangle}}^{\langle a \rangle},\ldots,{\cal D}_{N_{\langle a \rangle}-2}^{\langle a \rangle}$ 
cover the non-compactified part of the space outside star $a$; they are also
called {\em shells}. The inner boundary of domain ${\cal D}_{M_{\langle a \rangle}}^{\langle a \rangle}$
coincides exactly with the surface of star $a$. 
\item Domain ${\cal D}_{N_{\langle a \rangle}-1}^{\langle a \rangle}$ is the most external one; it extends
up to $r=+\infty$. We call this domain the {\em compactified external domain
(CED)} since thanks to some compactification it will be mapped to a finite
computational domain. 
\end{itemize}
Of course the two sets of domains overlap since 
${\cal D}_0^{\langle 1 \rangle} \cup \ldots \cup {\cal D}_{N_{\langle 1 \rangle}-1}^{\langle 1 \rangle} 
 = {\cal D}_0^{\langle 2 \rangle} \cup \ldots \cup {\cal D}_{N_{\langle 2 \rangle}-1}^{\langle 2 \rangle} = \Sigma_t$. 
The various domains are represented in Fig.~\ref{f:domains} for 
$N_{\langle 1 \rangle} = N_{\langle 2 \rangle} = 3$. 

Following the technique introduced previously \cite{BonazGM98a},
we define in each domain the {\em computational coordinates}
$(\xi,\theta',\varphi')$ according to\footnote{For the sake of
clarity we omit here the star 
indices ${\langle a \rangle}$ on the spherical coordinates 
$(r_{\langle a \rangle},\theta_{\langle a \rangle},\varphi_{\langle a \rangle})$
centered on star $a$.}
\be
	\theta' = \theta \ , \qquad \varphi' = \varphi
\ee
and
\begin{itemize}
\item in the nucleus:
\be \label{e:map_r_xi_1}
	r = \alpha_0 \left[ \xi + (3\xi^4 - 2\xi^6) F_0(\theta,\varphi)
		+ {1\ov 2} \left( 5\xi^3 - 3 \xi^5 \right)
		  G_0(\theta,\varphi) \right] \ ,  \qquad \xi \in [0,1] \ ; 
\ee
\item in the shells ($1\leq l \leq N_{\langle a \rangle}-2$) :
\be \label{e:map_r_xi_2}
	r = \alpha_l \left[ \xi + {1\over 4} \left( \xi^3 - 3\xi +2 \right)
		F_l(\theta,\varphi)
		+ {1\over 4} \left( -\xi^3 + 3\xi +2 \right)
		  G_l(\theta,\varphi) \right] + \beta_l
					\ , \qquad \xi \in [-1,1] \ ; 
\ee
\item in the CED:
\be \label{e:map_r_xi_3}
	r = {2 R_{\rm CED} \over 1 - \xi } \ , \qquad \xi \in [-1,1] \ .   
\ee
\end{itemize}
In the above relations, $\alpha_l$ and $\beta_l$ are some constants, the
functions $F_l(\theta,\varphi)$ and $G_l(\theta,\varphi)$ define the 
boundary of each domain: the outer boundary of the nucleus 
corresponds to $\xi=1$ and is given by the equation
\be
	r = \alpha_0 \left[ 1 + F_0(\theta,\varphi) 
			      + G_0(\theta,\varphi) \right] \ , 
\ee
where $F_0(\theta,\varphi)$ contains only odd Fourier harmonics in 
$\varphi$ and $G_0(\theta,\varphi)$ only even harmonics, 
the inner boundary of the shell no. $l$ ($1\leq l \leq N_{\langle a \rangle}-2$) 
corresponds to $\xi=-1$ and is given by the equation
\be
	r = \alpha_l \left[ -1 + F_l(\theta,\varphi) \right] + \beta_l \ ,
\ee
whereas its outer boundary corresponds to $\xi=1$ and is given by the equation
\be
	r = \alpha_l \left[ 1 + G_l(\theta,\varphi) \right] + \beta_l \ . 
\ee
Finally $R_{\rm CED}$ is the radius of the inner boundary of the CED, which
is assumed to be spherical. 

\subsection{Multi-domain spectral method} \label{s:spectr}

In each domain, we expand the various physical fields in a series 
of basis functions of $\xi$, $\theta'$ and $\varphi'$. We use Chebyshev 
polynomials in $\xi$, trigonometrical polynomials or associated Legendre
functions in $\theta'$, and Fourier series in $\varphi'$. 
The interested reader is referred to Sect.~III.A of Ref.~\cite{BonazGM98a} 
for more details about these spectral expansions. 
Let us denote by $N_r^{\langle a \rangle}(l)$ the number of coefficient in the $\xi$
expansion used in domain ${\cal D}_l^{\langle a \rangle}$, by $N_\theta^{\langle a \rangle}(l)$ 
the number of coefficients in the $\theta'$ expansion and by  
$N_\varphi^{\langle a \rangle}(l)$ the number of coefficients in the $\varphi'$ expansion. 
We employ a {\em collocation spectral method}, which means that in each
domain, a function can be described either by the coefficients of its
spectral expansion or by its value at some particular grid points, called
the {\em collocation points} \cite{CanutHQZ88}. The grids plotted in 
Fig.~\ref{f:domains} show actually these collocation points. 

The spectral method amounts to reducing linear
partial differential equations into a system of algebraic
equations for the coefficients of the spectral expansions.
We refer to \cite{BonazGM99b,GrandBGM00} for the details of this
multi-domain spectral method and here simply recall some basic features:
\begin{itemize}
\item As explained above, spherical-type coordinates $(\xi,\theta',\varphi')$
centered on each star are used: this ensures
a much better description of the stars  than by means of Cartesian coordinates. 
\item These spherical-type coordinates are surface-fitted coordinates: i.e.
the surface of each star lies at a constant value of the coordinate
$\xi$ thanks to the mapping $(\xi,\theta',\varphi')\mapsto (r,\theta,\varphi)$
defined by Eqs.~(\ref{e:map_r_xi_1})-(\ref{e:map_r_xi_2}). This 
ensures that the spectral method applied in each domain is free from any
Gibbs phenomenon (spurious oscillations generated by discontinuities).  
\item The outermost domain extends up to spatial infinity, thanks to the
mapping (\ref{e:map_r_xi_3}). 
This enables us to put exact boundary conditions
on the elliptic equations (\ref{e:nu})-(\ref{e:shift}) for the metric 
coefficients: spatial infinity
is the only location where the metric is known in advance (Minkowski metric). 
\item Thanks to the use of a spectral method 
in each domain, the numerical
error is {\em evanescent} for analytical fields (e.g. density fields for a 
$\gamma = 2$ equation of state), 
i.e. it decreases exponentially with the number
of coefficients (or equivalently collocation grid points) 
used in the spectral expansions
\cite{BonazGM99b,GrandBGM00}. 
\end{itemize}

\subsection{Splitting of the metric quantities}

Having introduced two sets of computational domains (grids), we linearly
split the metric potentials $\nu$, $\beta$ and $N^i$ into
\be \label{e:split_nu}
	\nu \ = \ \nu_{\langle 1 \rangle} + \nu_{\langle 2 \rangle} 
	    \ = \  \nu_{\langle 1 \rangle} + \nu_{\langle 2 \rightarrow 1 \rangle}
	    \ = \  \nu_{\langle 1 \rightarrow 2 \rangle} + \nu_{\langle 2 \rangle}  \ , 
\ee
\be \label{e:split_beta}
	\beta \ = \  \beta_{\langle 1 \rangle} + \beta_{\langle 2 \rangle} 
	    \ = \  \beta_{\langle 1 \rangle} + \beta_{\langle 2 \rightarrow 1 \rangle}
	    \ = \  \beta_{\langle 1 \rightarrow 2 \rangle} + \beta_{\langle 2 \rangle} \ , 
\ee
\be \label{e:split_shift}
	N^i \ = \  N^i_{\langle 1 \rangle} + N^i_{\langle 2 \rangle} 
	    \ = \  N^i_{\langle 1 \rangle} + N^i_{\langle 2 \rightarrow 1 \rangle}
	    \ = \  N^i_{\langle 1 \rightarrow 2 \rangle} + N^i_{\langle 2 \rangle} \ , 
\ee
where the quantities labeled by ``$\langle a \rangle$'' or 
``$\langle b \rightarrow a \rangle$'' 
($a=1,2$, $b=3-a$) are defined at the collocation
points of the domains ${\cal D}_l^{\langle a \rangle}$ centered on star $a$,
and the quantities labeled by ``$\langle a \rangle$'' and ``$\langle a \rightarrow b \rangle$''
represents the same physical field but described at different collocation
points (those of domain sets ${\cal D}_l^{\langle a \rangle}$ and ${\cal D}_l^{\langle b \rangle}$
respectively), i.e. $\nu_{\langle 1 \rightarrow 2 \rangle} = \nu_{\langle 1 \rangle}$,
$\nu_{\langle 2 \rightarrow 1 \rangle} = \nu_{\langle 2 \rangle}$, etc... 

The basic idea underlying the splittings 
(\ref{e:split_nu})-(\ref{e:split_shift}) is that for each metric potential, 
there are two primary quantities, those labeled by ``$\langle 1 \rangle$'' and
``$\langle 2 \rangle$'', which are ``mostly generated'' by respectively star 1 and
star 2 and which we called the {\em auto-potentials}
[the precise definitions are given by Eqs.~(\ref{e:nu_a})-(\ref{e:shift_a})
below]. 
The auto-potentials are obtained by solving the gravitational field
equations, on domains ${\cal D}_l^{\langle 1 \rangle}$ for the ``$\langle 1 \rangle$'' potentials,
and on ${\cal D}_l^{\langle 2 \rangle}$ for the ``$\langle 2 \rangle$'' ones. The quantities
labeled by ``$\langle 1 \rightarrow 2 \rangle$'' [resp. ``$\langle 2 \rightarrow 1 \rangle$''] 
are then merely representations of the ``$\langle 1 \rangle$'' 
[resp. ``$\langle 2 \rangle$''] 
auto-potentials at the collocation points associated with the companion
star. For this reason, we shall call them the {\em comp-potentials}. 

Following the splittings (\ref{e:split_nu})-(\ref{e:split_shift}),
the gravitational field equations (\ref{e:nu})-(\ref{e:shift}) are themselves 
split in two parts.
$\nu_{\langle 1 \rangle}$ and $\nu_{\langle 2 \rangle}$ are thus defined as the solutions of the 
two equations
\be \label{e:nu_a}
	\ulap \nu_{\langle a \rangle} = 4 \pi A^2 (E_{\langle a \rangle} + S_{\langle a \rangle}) + 
			Q_{\langle a \rangle} + Q_{\langle b \rightarrow a \rangle} 
		- \unab_i \nu_{\langle a \rangle} \left[
			\unab^i \beta_{\langle a \rangle} +
			(\unab^i \beta)_{\langle b \rightarrow a \rangle} \right] \ ,
		\qquad  a=1,2 \quad (b=3-a) \ ,
\ee
whereas $\beta_{\langle 1 \rangle}$ and $\beta_{\langle 2 \rangle}$ are
defined as the solutions of the two equations
\begin{eqnarray} 
\ulap \beta_{\langle a \rangle} & = & 4 \pi A^2 S_{\langle a \rangle} + 
		{3 \over 4} ( Q_{\langle a \rangle} + Q_{\langle b \rightarrow a \rangle} )
        - {1 \over 2} \unab_i \nu_{\langle a \rangle} \left[
			\unab^i \nu_{\langle a \rangle} +
			(\unab^i \nu)_{\langle b \rightarrow a \rangle} \right] \nonumber \\
		& & - {1 \over 2} \unab_i \beta_{\langle a \rangle} \left[
			\unab^i \beta_{\langle a \rangle} +
			(\unab^i \beta)_{\langle b \rightarrow a \rangle} \right]  \ , 
			\qquad  a=1,2 \quad (b=3-a) \ ,	 \label{e:beta_a}
\end{eqnarray}
and $N^i_{\langle 1 \rangle}$ and $N^i_{\langle 2 \rangle}$ are defined as the solutions of the 
two equations
\begin{eqnarray} 
\ulap N^i_{\langle a \rangle} + {1 \over 3} \unab^i \left( \unab_j N^j_{\langle a \rangle} \right) & = & 
        - 16 \pi N A^2 (E_{\langle a \rangle} + p_{\langle a \rangle}) U_{\langle a \rangle}^i 
	+ N {\tilde K}^{ij}_{\langle a \rangle} \Big(
		6 \left[ \unab_j \beta_{\langle a \rangle} +
			(\unab_j \beta)_{\langle b \rightarrow a \rangle} \right] 
					\nonumber \\
 & & - 8 \left[ \unab_j \nu_{\langle a \rangle} +
			(\unab_j \nu)_{\langle b \rightarrow a \rangle} \right]   
	\Big) \ , \qquad  a=1,2 \quad (b=3-a) \ .  \label{e:shift_a}
\end{eqnarray}
In these equations, $E_{\langle a \rangle}$, $S_{\langle a \rangle}$, $p_{\langle a \rangle}$, $U_{\langle a \rangle}^i$
are the quantities relative to the fluid of star $a$ only and
defined respectively by Eqs.~(\ref{e:ener_euler}),
(\ref{e:s_euler}), (\ref{e:stress-energy}) and (\ref{e:def_U}).
${\tilde K}^{ij}_{\langle a \rangle}$ is defined from $N^i_{\langle a \rangle}$ according to 
\be \label{e:tkij_def}
	{\tilde K}^{ij}_{\langle a \rangle} :=  - {1 \over 2 N}
          \left\{
                \unab^i N^j_{\langle a \rangle} + \unab^j N^i_{\langle a \rangle}
                  - {2 \over 3} f^{ij} \unab_k N^k_{\langle a \rangle} 
          \right\} \ , \qquad  a=1,2 \ , 
\ee
so that the total extrinsic curvature is given by 
$K^{ij} = ({\tilde K}^{ij}_{\langle 1 \rangle} + {\tilde K}^{ij}_{\langle 2 \rangle})/A^2$. 
Finally $Q_{\langle a \rangle}$ and $Q_{\langle b \rightarrow a \rangle}$ are defined by
\be
	Q_{\langle a \rangle} := A^2 f_{ik} f_{jl} {\tilde K}^{kl}_{\langle a \rangle} 
		    {\tilde K}^{ij}_{\langle a \rangle} \ ,  \qquad  a=1,2 \ ,
\ee
\be
	Q_{\langle b \rightarrow a \rangle} := A^2 f_{ik} f_{jl} {\tilde K}^{kl}_{\langle a \rangle} 
		    {\tilde K}^{ij}_{\langle b \rightarrow a \rangle} \ , 
				\qquad  a=1,2 \quad (b=3-a) \ , 
\ee
where ${\tilde K}^{ij}_{\langle b \rightarrow a \rangle}$ is the same physical field 
than ${\tilde K}^{ij}_{\langle b \rangle}$ but numerically described at the collocation
points of the domains ${\cal D}_l^{\langle a \rangle}$, ${\tilde K}^{ij}_{\langle b \rangle}$ being
given at the collocation points of the domains ${\cal D}_l^{\langle b \rangle}$. 

It is straightforward to check that adding the two equations (\ref{e:nu_a})
results in Eq.~(\ref{e:nu}), adding the two equations (\ref{e:beta_a})
results in Eq.~(\ref{e:beta}) and adding the two equations (\ref{e:shift_a})
results in Eq.~(\ref{e:shift}). Therefore, having obtained solutions
$\nu_{\langle a \rangle}$, $\beta_{\langle a \rangle}$ and $N^i_{\langle a \rangle}$ of the equations 
(\ref{e:nu_a})-(\ref{e:shift_a}), we can form the solution of the gravitational
field equations (\ref{e:nu}), (\ref{e:beta}) and (\ref{e:shift})
via Eqs.~(\ref{e:split_nu})-(\ref{e:split_shift}). 

The advantage of solving the system of $2\times 5=10$ PDEs 
(\ref{e:nu_a})-(\ref{e:shift_a}), instead of solving the system of 5 PDEs
(\ref{e:nu})-(\ref{e:shift}), is that the source terms (right-hand-side) of
the former are mostly concentrated on one of the two stars and therefore
well described by one of the two domain sets introduced in 
Sect.~\ref{s:domains}. This is not true for the source terms involving the
comp-potentials
``$\langle b \rightarrow a \rangle$''. However
these terms enter only via quadratic combinations
in which each of them is multiplied by the gradient of an auto-potential
term, which is small where the comp-potential is large, so that the
product of the two is smaller than the other sources terms, such as
the scalar product of gradients of auto-potentials. The same 
considerations hold for $Q_{\langle b \rightarrow a \rangle}$ which appears to be
much smaller than  $Q_{\langle a \rangle}$.
According to these remarks, Eq.~(\ref{e:nu_a}) for $\nu_{\langle 1 \rangle}$ is
naturally solved on domains ${\cal D}_l^{\langle 1 \rangle}$, 
Eq.~(\ref{e:nu_a}) for $\nu_{\langle 2 \rangle}$ is solved on domains 
${\cal D}_l^{\langle 2 \rangle}$, and more generally, each equation for an auto-potential
is solved onto the domains set centered on the corresponding star. 

Once the auto-potentials are known (at a given step of the iterative
procedure described in the next section), there remains to compute
the corresponding comp-potentials. This means that given e.g. $\nu_{\langle 1 \rangle}$
at the collocation points of domains ${\cal D}_l^{\langle 1 \rangle}$, one has to 
compute its values $\nu_{\langle 1 \rightarrow 2 \rangle}$ at the collocation points
of domains ${\cal D}_l^{\langle 2 \rangle}$. One may think first 
to use some interpolation technique since the two sets of domains overlap.
But this will necessarily introduce some ``numerical noise''. 
We will proceed differently, taking advantage of the use of a spectral
method. Indeed, the values of the field $\nu_{\langle 1 \rangle}$ at the collocation
points of domains ${\cal D}_l^{\langle 1 \rangle}$ is not the only 
numerical representation of $\nu_{\langle 1 \rangle}$ we have at our disposal. 
We can use the alternative representation by the set of coefficients
of its spectral expansion in each domain ${\cal D}_l^{\langle 1 \rangle}$
($0\leq l \leq N_{\langle 1 \rangle}-1$) (cf. Sect.~\ref{s:spectr}). By means of this
spectral expansion, we can compute the value of $\nu_{\langle 1 \rangle}$ at 
{\em any}
point in the domain ${\cal D}_l^{\langle 1 \rangle}$, not necessarily a collocation point. 
Hence, given a collocation point $(\xi_i,\theta'_j,\varphi'_k)$ of
domain ${\cal D}_{l_0}^{\langle 2 \rangle}$, we first compute the 
corresponding physical spherical coordinates 
$(r_{\langle 2 \rangle},\theta_{\langle 2 \rangle},\varphi_{\langle 2 \rangle})$ via 
Eqs.~(\ref{e:map_r_xi_1})-(\ref{e:map_r_xi_3}), then the corresponding
Cartesian coordinates $(x_{\langle 2 \rangle},y_{\langle 2 \rangle},z_{\langle 2 \rangle})$; these latter are
translated into Cartesian coordinates $(x_{\langle 1 \rangle},y_{\langle 1 \rangle},z_{\langle 1 \rangle})$
via Eq.~(\ref{e:x-X}). We finally obtain the
corresponding spherical coordinates $(r_{\langle 1 \rangle},\theta_{\langle 1 \rangle},\varphi_{\langle 1 \rangle})$ 
centered on star 1. We then determine in which domain ${\cal D}_l^{\langle 1 \rangle}$
this point is localized and to which value of the coordinate $\xi$
it corresponds by inverting the relations 
(\ref{e:map_r_xi_1})-(\ref{e:map_r_xi_3}). Then we may use the spectral
expansion of $\nu_{\langle 1 \rangle}$ to get the searched value:
\begin{eqnarray}
  \nu_{\langle 1 \rightarrow 2 \rangle}(l_0,\xi_i,\theta'_j,\varphi'_k) & = &  
	\sum_{k=0}^{N_\varphi^{\langle 1 \rangle}(l)-1} \Bigg[
	\sum_{j=0}^{N_\theta^{\langle 1 \rangle}(l)-1} \left(
	\sum_{i=0}^{N_r^{\langle 1 \rangle}(l)-1} \hat \nu_{lkji}\, X_{kji}(\xi) \right) \, 
  \Theta_{kj}(\theta_{\langle 1 \rangle})	\Bigg]
	\Phi_k(\varphi_{\langle 1 \rangle})	\ ,	\label{e:interpol}
\end{eqnarray}
where the $\hat \nu_{lkji}$ are the coefficients of  $\nu_{\langle 1 \rangle}$ in domain
${\cal D}_l^{\langle 1 \rangle}$, $X_{kji}$ denotes the basis functions in $\xi$ 
(typically Chebyshev polynomials), $\Theta_{kj}$ the basis functions
in $\theta$ (typically $\cos n\theta$ or $\sin n\theta$) and 
$\Phi_k$ the basis functions in $\varphi$ (Fourier series). These functions
depend on the type of domain (nucleus, shell or CED) and are described in
details in Sect.~III.A of Ref.~\cite{BonazGM98a}.

\subsection{Iterative procedure} \label{s:iter}

Within our procedure, 
a quasiequilibrium binary neutron star configuration is obtained by 
specifying:
\begin{enumerate}
\item the equation of state (\ref{e:eos}) for each star;
\item the rotation state: either rigidly rotating (synchronized binaries,
Sect.~\ref{s:rigid}) or irrotational flow (Sect.~\ref{s:irrot});
\item the coordinate distance $d:=|X_{\langle 2 \rangle}-X_{\langle 1 \rangle}|$ between the two stellar
centers;
\item \label{s:central_dens} the central enthalpies 
$H_{\langle 1 \rangle}^{\rm c}$ and $H_{\langle 2 \rangle}^{\rm c}$ in each 
star, or equivalently, via (\ref{e:eos}),
the central density in each star 
(with our definition of the stellar
center, this coincides with the maximum density). 
\end{enumerate}
As we discuss below, item \ref{s:central_dens} can be replaced by the
specification of the baryon mass of each star. 

\subsubsection{Initial conditions}

The above parameters being set, we start by computing initial conditions
for the iterative procedure. These initial conditions are constituted 
by two numerical solutions for spherically symmetric static isolated neutron
stars, of respective central enthalpy $H_{\langle 1 \rangle}^{\rm c}$ and $H_{\langle 2 \rangle}^{\rm c}$. 
$M_{\langle 1 \rangle}$ and $M_{\langle 2 \rangle}$ being the gravitational masses of
these spherical symmetric models, we set the $X$ coordinates of the two
stellar centers according to the Newtonian-like formulas:
\be \label{e:positions}
	X_{\langle 1 \rangle} = - {M_{\langle 2 \rangle} \over M_{\langle 1 \rangle} + M_{\langle 2 \rangle}}
			\, d 
	\qquad \mbox{and} \qquad 
	X_{\langle 2 \rangle} =  {M_{\langle 1 \rangle} \over M_{\langle 1 \rangle} + M_{\langle 2 \rangle}}
			\, d  \ .
\ee
These coordinates will remain fixed during the iteration. Only the location
of the rotation axis $X_{\rm rot}$, initially set to $0$, will change
(see Fig.~\ref{f:coord}). Accordingly the formul\ae\ (\ref{e:positions})
have no physical meaning whatsoever. They can be viewed as the setting of
the origin of the coordinate system $(X,Y,Z)$. The location of this origin
is a priori arbitrary, only the distance $d$ between the two stellar centers
having a physical meaning; the setting (\ref{e:positions}) simply insures
that this origin is not too far from the rotation axis. 

The angular velocity $\Omega$ is initialized according to a formula
for second order Post-Newtonian spherical stars \cite{BlancDIWW95,Kidde95}:
\begin{eqnarray} 
	 \Omega_{\rm ini}^2   & = &  
	{ M_{\rm ini} \over d^3 } \Bigg\{ 1 - { M_{\rm ini} \over d }	
	\left[ {11\over 4} +  {2 R^2  \over d^2} \, \gamma
		- {12 \over 25} {R^4  \over d^4} \, \gamma^2 \right] 
			\nonumber \\
  & & \ \qquad \quad + \left( { M_{\rm ini} \over d } \right) ^2
	\left[ {69\over 8} + {11\over 4} {R^2 \over d^2} 
		\gamma + {17 \over 25}  {R^4 \over d^4} \gamma^2 \right] 
		\Bigg\} \ ,  			\label{e:omega_ini}
\end{eqnarray}
where $M_{\rm ini} := M_{\langle 1 \rangle} + M_{\langle 2 \rangle}$,
$R$ is the coordinate radius of one of the two stars\footnote{
Equation (\ref{e:omega_ini}) is valid only for equal-mass stars binaries. 
There also exists a more complicated formula for stars with different masses.}
(which is spherical initially)
and $\gamma = \gamma_{\rm irrot} := 0$ for irrotational binaries, whereas 
$\gamma = \gamma_{\rm corot} := 5 I_{\langle a \rangle} / (2 M_{\langle a \rangle} R_{\langle a \rangle}^2)$ 
for corotating binaries, $I_{\langle a \rangle}$ being
the moment of inertia of star $a$. For this last quantity,
we use as an ansatz the exact value for a Newtonian $n=1$
polytrope, which results in $\gamma_{\rm corot} = 5/3\, (1-6/\pi^2)$,
independent of $a$. 

The metric auto-potentials are initialized as follows: $\nu_{\langle a \rangle}$
and $\beta_{\langle a \rangle}$ are set to the values of $\nu$ and $\beta$ for the
static spherical models. The shift $N^i_{\langle a \rangle}$ is initialized to 
the first-order Post-Newtonian value for spherical incompressible binaries
(this value can be obtained by taking the limit for a spherical star 
of the equations presented in Ref. \cite{Tanig99}) 
\be
	N^i_{\langle a \rangle} = {7\over 8} W^i_{\langle a \rangle} 
		- {1\over 8} \left( \unab^i \chi_{\langle a \rangle} + 
		 \unab^i W^j_{\langle a \rangle} x_j \right) \ , \qquad  a=1,2 \ , 	
\ee
with 
\be
	W^X_{\langle a \rangle} = 0 \ ; \qquad
	W^Y_{\langle a \rangle} = \left\{ \begin{array}{lr}
	 \epsilon_{\langle a \rangle} {6 M_{\langle a \rangle} \Omega_{\rm ini} d \over 
		\left(1+ {M_{\langle a \rangle} / M_{\langle b \rangle}} \right)
		  R_{\langle a \rangle}}
	\left( 1 - {r_{\langle a \rangle}^2 \over 3 R_{\langle a \rangle}^2} \right) & 
	\qquad \mbox{for} \quad r_{\langle a \rangle} \leq  R_{\langle a \rangle} 			\\
	\epsilon_{\langle a \rangle} {4 M_{\langle a \rangle} \Omega_{\rm ini} d \over 
	\left(1+ {M_{\langle a \rangle} / M_{\langle b \rangle}} \right) r_{\langle a \rangle}} &
	\qquad \mbox{for} \quad r_{\langle a \rangle} >  R_{\langle a \rangle} 			
				\end{array} \right. 
	\ ; \qquad
	W^Z_{\langle a \rangle} = 0  \ , 
\ee
($ \epsilon_{\langle 1 \rangle} := -1$, $ \epsilon_{\langle 2 \rangle} := 1$, $b=3-a$) and
\be
	\chi_{\langle a \rangle} = \left\{ \begin{array}{lr}
	 {2 M_{\langle a \rangle} \Omega_{\rm ini} d \over 
		\left(1+ {M_{\langle a \rangle} / M_{\langle b \rangle}} \right)
		  R_{\langle a \rangle}} \, y_{\langle a \rangle} \, 
	\left( 1 - {3 r_{\langle a \rangle}^2 \over 5 R_{\langle a \rangle}^2} \right) & 
	\qquad \mbox{for} \quad r_{\langle a \rangle} \leq  R_{\langle a \rangle} 			\\
	{4 M_{\langle a \rangle} \Omega_{\rm ini} d R_{\langle a \rangle}^2 \over 5  
	\left(1+ {M_{\langle a \rangle} / M_{\langle b \rangle}} \right) } 
		\, {y_{\langle a \rangle} \over r_{\langle a \rangle}^3} &
	\qquad \mbox{for} \quad r_{\langle a \rangle} >  R_{\langle a \rangle} 			
				\end{array} \right.
\ee
We refer to Sect.~\ref{s:domains} for the definition of the
coordinates $X$, $Y$, $Z$, $r_{\langle a \rangle}$ and $y_{\langle a \rangle}$ involved in these
formul\ae. It appeared that the initial shift vector given above results in 
a too large angular velocity in the first steps. Therefore, we artificially
lower it by multiplying it by $0.6$. 
From this initial value of the shift, we get initial values of 
${\tilde K}^{ij}_{\langle a \rangle}$ via Eq.~(\ref{e:tkij_def}), and 
initial values of 
$\mib{B}$ and $\mib{U}_0$ via Eqs.~(\ref{e:shift_non_rot}) and
(\ref{e:U_0}).  

Regarding the fluid quantities, the 3-velocity $\mib{U}$ is initialized
to $\mib{U}_0$ in the synchronized case, whereas in the irrotational
case, $\Psi_0$ is initialized to zero and $\Psi$ is initialized accordingly
via Eq.~(\ref{e:Psi_Psi0}); the Lorentz factor $\Gamma_{\rm n}$ is then
initialized via Eq.~(\ref{e:gamma_n_flat}) and the 3-velocity $\mib{U}$
is initialized according to Eq.~(\ref{e:U_flat_grad_Psi}). We get then
initial values of the Eulerian energy density $E$ and the trace of stress 
tensor $S$ via Eqs.~(\ref{e:ener_euler}) and (\ref{e:s_euler}). In these
equations, we use for the proper energy density $e$ and pressure $p$ the
values of the spherically symmetric initial stellar models. 

\subsubsection{Description of one step} \label{s:descrip_step}

At a given step, we start by determining the value of the orbital 
angular velocity $\Omega$ and the value of the $X$ coordinate of the
rotation axis, $X_{\rm rot}$ (see Fig.~\ref{f:coord}), by taking the gradient
along $X$ of the first integral of motion (\ref{e:int_prem_3p1}).
Demanding that the enthalpy $H$ be maximal at the center of each star
(our definition of {\em center}), this results in the two equations
\be \label{e:orbite}
	 \left. {\partial\over \partial X} \ln \Gamma_0 \right| _{(X_{\langle a \rangle},0,0)}
	=  \left. {\partial\over \partial X} (\nu + \ln \Gamma) 
	  \right| _{(X_{\langle a \rangle},0,0)}   \qquad a = 1,2\ , 
\ee
where $\ln \Gamma_0$ can be expressed in terms of $\Omega$ and $X_{\rm rot}$
thanks to Eqs.~(\ref{e:gamma_0}), (\ref{e:U_0}), (\ref{e:conf_plat}) and
(\ref{e:shift_non_rot}):
\be
	\ln\Gamma_0 = -{1\over 2} \ln \left\{
	1 - {A^2\over N^2} \left[ \left( \Omega Y + N^X \right) ^2 
		+ \left( \Omega(X-X_{\rm rot})-N^Y \right) ^2 
		+ \left( N^Z \right) ^2 \right] \right\} \ . 
\ee
Inserting this relation into Eq.~(\ref{e:orbite}) and setting $Y=Z=0$,
$X=X_{\langle 1 \rangle}$ or $X_{\langle 2 \rangle}$ results in a system of two equations for the
two unknowns $\Omega$ and $X_{\rm rot}$. This system is solved by standard
methods. 
Having determined $\Omega$ and $X_{\rm rot}$, we can compute the components
of the orbiting velocity $\mib{U}_0$, via Eqs.~(\ref{e:U_0}) and
(\ref{e:shift_non_rot}):
\be
	U_0^X = - {1\over N} \left( \Omega Y + N^X \right) ; \qquad
	U_0^Y = {1\over N} \left( \Omega(X-X_{\rm rot})-N^Y \right) ; \qquad
	U_0^Z = - {N^Z\over N} \ , 
\ee
where $N$, $N^X$, $N^Y$ and $N^Z$ are the values of the lapse function and
the components of the non-rotating-coordinates shift vector taken from the
previous step. From $\mib{U}_0$, we of course compute the Lorentz factor
$\Gamma_0$ by Eq.~(\ref{e:gamma_0}). 
The fluid 3-velocity with respect to the Eulerian observer, $\mib{U}$
is set to $\mib{U}_0$ in the synchronized case, where  
in the irrotational case, $\Psi$ is deduced from $\mib{U}_0$ and the
previous step value of $\Psi_0$
via Eq.~(\ref{e:Psi_Psi0}); the Lorentz factor $\Gamma_{\rm n}$ is then
computed via Eq.~(\ref{e:gamma_n_flat}) and the 3-velocity $\mib{U}$
follows from Eq.~(\ref{e:U_flat_grad_Psi}). 
The Lorentz factor $\Gamma$ between the fluid and the co-orbiting 
observer is deduced from the above quantities via Eq.~(\ref{e:gamma_flat}). 

The elliptic equation (\ref{e:psinum2}) for $\Psi_0$ is then solved by the
numerical method described in Appendix~\ref{s:app_psi}. 

We then adapt the computational domains to the stars as follows. The
first integral of motion (\ref{e:int_prem_3p1}) is written, 
following the splitting (\ref{e:split_nu})
\be \label{e:int_prem_final}
	H = H_{\langle a \rangle}^{\rm c} + \nu_{\langle a \rangle}^{\rm c} + \Phi_{\langle a \rangle,\rm ext}^{\rm c}
		-  \nu_{\langle a \rangle} - \Phi_{\langle a \rangle,\rm ext} \ ,
\ee
where we have introduced the ``external'' potential
\be
    \Phi_{\langle a \rangle,\rm ext} := \nu_{\langle b \rightarrow a \rangle}
		-  \ln \Gamma_0 + \ln \Gamma
\ee
and the superscript ``${\rm c}$'' stands for values at the center of
the star. First, we rescale the auto-potential $\nu_{\langle a \rangle}$ by a factor
$\alpha^2$ to make sure that the enthalpy vanishes at the point
$\theta_{\langle a \rangle}=\pi/2$, $\varphi_{\langle a \rangle}=0$ on the external boundary
of domain ${\cal D}_{M_{\langle a \rangle}-1}^{\langle a \rangle}$:
\be 
	\alpha^2 = { H_{\langle a \rangle}^{\rm c} 
	+ \Phi_{\langle a \rangle,\rm ext}^{\rm c} - \Phi_{\langle a \rangle,\rm ext}^{\rm s}
	\over 
	\nu_{\langle a \rangle}^{\rm s} - \nu_{\langle a \rangle}^{\rm c}	} \ , 
\ee
where the superscript ``${\rm s}$'' stands for values at the point
$\xi=1$, $\theta_{\langle a \rangle}=\pi/2$, $\varphi_{\langle a \rangle}=0$ of domain 
${\cal D}_{M_{\langle a \rangle}-1}^{\langle a \rangle}$.
When the iteration converges, $\alpha$ tends to 1. 
We then replace $\nu_{\langle a \rangle}$ by $\alpha^2 \nu_{\langle a \rangle}$ in 
Eq.~(\ref{e:int_prem_final}) to get the enthalpy field in all space.
Following the technique described in Ref.~\cite{BonazGM98a}, 
we then compute new functions $F_l(\theta,\varphi)$ and $G_l(\theta,\varphi)$
in the mappings (\ref{e:map_r_xi_1}) and (\ref{e:map_r_xi_2}) in order
to make the outer boundary of domain ${\cal D}_{M_{\langle a \rangle}-1}^{\langle a \rangle}$ coincide
exactly with the surface of the star. Since the
collocation points of the new mapping do not coincide (in the physical space,
described by the coordinates $(r_{\langle a \rangle},\theta_{\langle a \rangle},\varphi_{\langle a \rangle})$) with
that of the previous mapping, the values of the enthalpy field at the
new collocation points have to be computed. The details of this computations
are explained in Sect.~V.A of Ref.~\cite{BonazGM98a}

From this new value of $H$, we compute the fluid proper baryon density
$n$, proper energy density $e$ and pressure $p$ via the EOS (\ref{e:eos}). 
We then get the Eulerian energy density $E$ and the trace of stress 
tensor $S$ via Eqs.~(\ref{e:ener_euler}) and (\ref{e:s_euler}). 
These last quantities are subsequently used to evaluate the source
terms of the elliptic equations (\ref{e:nu_a})-(\ref{e:shift_a}) for the 
gravitational potentials. These equations are solved
by means of the multi-domain scalar and vector Poisson solvers
for non-compact sources described in details in 
Refs.~\cite{BonazGM98a,GrandBGM00}. 
In particular the vector Poisson equation
(\ref{e:shift_a}) for the auto shift $N^i_{\langle a \rangle}$ 
is reduced to a set of 4 scalar Poisson equations according to the  
scheme used by Shibata and Oohara \cite{ShibaO97,Shiba99c}. 

Before the beginning of a new step, some relaxation is performed onto the
enthalpy field and the auto-potentials, according to 
\be
  Q^J \leftarrow  \lambda  Q^J + (1-\lambda) Q^{J-1}  \ , 
\ee
where $Q$ stands for any of the fields $H$, $\nu_{\langle a \rangle}$,
$\beta_{\langle a \rangle}$ and $N^i_{\langle a \rangle}$ ($a=1,2$), 
$J$ (resp. $J-1$) labels the current step (resp. previous step), 
and $\lambda$ is the relaxation factor, typically chosen to be $0.5$ for
$H$ and $0.65$ for the auto-potentials. 

For appreciably relativistic configurations, it appeared that the 
above relaxation is not sufficient to ensure the convergence. 
In this case, we update the comp-potentials not every step but every $m$
steps, with typically $m=8$. This slows the convergence but enforces
it.

\subsubsection{Convergence to a given baryon mass} \label{s:conv_bar_mass}

In order to compute evolutionary sequences of binary neutron stars, one should
be able to compute configurations for a given baryon mass, since this
quantity is conserved during the gravitational-radiation driven evolution of
the system. The baryon mass, given by Eq.~(\ref{e:m_bar}),
is not a natural parameter we can set in our procedure. 
As stated above, the freely specifiable
parameters which fix one configuration are the coordinate distance $d$
between the two stellar centers and the central enthalpies 
$H_{\langle 1 \rangle}^{\rm c}$ and $H_{\langle 2 \rangle}^{\rm c}$ in each 
star. However, we can use the iteration procedure itself to make the
final configuration have a specified baryon mass. Indeed, since the 
baryon mass is an increasing function of the central enthalpy (at least
for the stable stars we are studying), we multiply at each step, the central
enthalpy $H_{\langle 1 \rangle}^{\rm c}$ by the factor
\be
	\eta := \left( {2 + \zeta \over 2 + 2\zeta} \right) ^{1\over 4} \ , 
\ee
where $\zeta$ is the relative discrepancy between the actual baryon mass
at the considered step, $M_{\rm B}^{\langle 1 \rangle\, J}$ and the
wanted baryon mass $M_{\rm B}^{\langle 1 \rangle}$:
$\zeta := M_{\rm B}^{\langle 1 \rangle\, J} / M_{\rm B}^{\langle 1 \rangle}
-1$. When the iterative procedure converge, the factor $\eta$ tends to one. 
The same treatment is performed on star 2. 

\subsubsection{End of the iteration}

To control the convergence of the iterative procedure, we introduce the
relative difference between the enthalpy fields of two successive steps:
\be \label{e:delta_H}
	\delta H := {\sum_i \left| H^J(x_i) - H^{J-1}(x_i) \right| \over 
			\sum_i \left| H^{J-1}(x_i) \right| } \ , 
\ee
where the summation is extended to all the collocation points $x_i$ inside
the star and $J$ is the step label. 

We use typically $\delta H = 10^{-7}$ as a criterion to end the iteration. 
For very high precision calculations (check with analytical solutions,
see below), we use instead $\delta H = 10^{-12}$. 

\subsection{Treatment of cusps} \label{s:cusp}

For very close configurations, an angular point (cusp) may appear at the 
surface of the stars, similar to that in the Roche lobe at the Lagrange
point $L_1$ in the Roche problem. At this point the enthalpy gradient
$\partial H/\partial r$ vanishes in the direction of the companion. 
The surface of the star is then no longer smooth and
surface cannot be decribed by the differentiable
mapping (\ref{e:map_r_xi_1})-(\ref{e:map_r_xi_2}), because the 
functions $F_l(\theta,\varphi)$ and $G_l(\theta,\varphi)$ are assumed to 
be expandable in $\cos(k\theta)$ or $\sin(k\theta)$ series and in Fourier
series in $\varphi$, 
which implies that they are smooth functions of $(\theta,\varphi)$. 

The solution to this problem consists in freezing the adaptation 
of the mapping to the stellar surface when the enthalpy gradient becomes
too small at the surface point which faces the companion star. More
precisely, we define the ratio
\be \chi := {(\partial H/\partial r)_{\rm eq,comp}\over
     (\partial H/\partial r)_{\rm pole} } \ , 
\ee
where the index ``${\rm eq,comp}$'' stands for the value at the point 
$(\theta_{\langle a \rangle}=\pi/2,\ \varphi_{\langle a \rangle}=0)$ on 
the stellar surface, whereas the ``index'' ${\rm pole}$ stands for the value at 
the point $(\theta_{\langle a \rangle}=0,\ \varphi_{\langle a \rangle}=0)$
on the stellar surface. 
When $\chi$ passes below a certain threshold $\chi_{\rm fr}$
during the iteration process, we stop the adaptation of the mapping
to the surface of the star. $\chi_{\rm fr}$ is chosen typically 
chosen between $0.3$ and $0.55$. 

In this case, a Gibbs phenomenon is present. The accuracy of the calculation
is then lower than when the mapping is adapted to the surface of the star. 
However, since the difference between the stellar surface at the
domain boundary is pretty small, the Gibbs phenomenon 
is rather limited.

For irrotational configurations, the non-coincidence of the stellar surface
with a domain boundary introduces a small error in the resolution of
the equation (\ref{e:psinum2}) for the velocity potential $\Psi_0$
by means of the technique explained in Appendix~\ref{s:app_psi}.

\subsection{Numerical implementation}

The numerical code implementing the method described above is
written in the {\sc LORENE} ({\sc Langage Objet pour la
RElativit\'e Num\'eriquE}) language 
\cite{MarckGGN00}, which is a C++ based language for numerical relativity. 
A typical run makes use of 6 domains 
($N_{\langle 1 \rangle} = N_{\langle 2 \rangle} = 3$ and
 $M_{\langle 1 \rangle} = M_{\langle 2 \rangle} = 1$), with
$N_r \times N_\theta \times N_\varphi = 33 \times 21 \times 20$ coefficients 
in each domain.  The corresponding memory requirement
is 232~MB for an irrotational configuration. 
A computation involves $\sim 250$ steps,
which takes 14~h on one CPU of a SGI Origin200 computer 
(MIPS~R10000 processor at 180 MHz).
If the number of coefficients is lowered to 
$N_r \times N_\theta \times N_\varphi = 25 \times 17 \times 16$,
the memory requirement and CPU times becomes respectively 
100~MB and 6~h~30~min. 

Note that due to the rather small memory requirement, 
runs can be performed in parallel on a multi-processor platform. 
This is especially useful to compute sequences of configurations.  

Both Newtonian and relativistic configurations, either corotating 
or irrotational, are calculated by the same code. 
Only the parts of the computation specific to one of these four
cases are treated by different branches of the code.

\section{Tests of the numerical code} \label{s:tests}

After constructing a numerical code for calculation of binary neutron stars,
we must assert its validity by performing self-consistency checks and
comparing the results with those of analytic solutions or
those of  previous numerical works.
The plan of the tests of the numerical code is as follows.
For the irrotational configurations:
\begin{enumerate}
  \item Check the convergence  of the iterative procedure.

  \item Check the convergence of the global quantities when increasing
	the number of coefficients of the spectral method.

  \item Check the decay of the relative error in the virial theorem
	for Newtonian binary systems when increasing the number of
	coefficients of the spectral method.

  \item Check the agreement with some analytic solutions for Newtonian
	binary systems.

  \item Check the agreement with previous numerical solutions
	for Newtonian binary systems.

  \item Check the coincidence of the results of the purely Newtonian
	calculation with those of general relativistic one
	with small compactness.
\end{enumerate}
and for the corotating configurations:
\begin{enumerate}
  \item Check the agreement with previous numerical solutions
	of relativistic binary systems.
\end{enumerate}
For the purpose of the test computations,
we consider identical star binary systems
with the polytropic equation of state
\begin{eqnarray}
 n(H) & = &
	\left[ {\gamma-1\over \gamma} {m_{\rm B}  \over \kappa} (\exp(H) - 1)
 	    \right] ^{1/(\gamma-1)} \label{e:eos_poly_n} \\
 p(H) & = & \kappa \,  n(H)^\gamma \label{e:eos_poly_p} \\
 e(H) & = & {\kappa \over \gamma-1} n(H)^\gamma + m_{\rm B}  \, n(H) \ ,
			\label{e:eos_poly_e} 
\end{eqnarray}
where $\gamma$, $\kappa$ and $m_{\rm B}$ are some constants. 
For $m_{\rm B}$ we will use $m_{\rm B} = 1.66 \times 10^{-27} {\ \rm kg}$
(mean baryon mass).

\subsection{Self-consistency checks}

\begin{figure}
\centerline{ \epsfig{figure=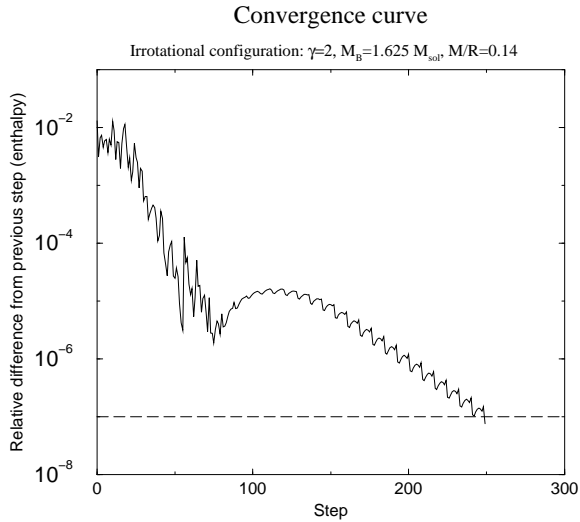,height=7cm} }
\caption[]{\label{f:conv_curve} 
Convergence (measured by the relative difference $\delta H$
in the enthalpy field between two successive steps) of the iterative 
procedure for one of the
irrotational models with 
$N_r \times N_\theta \times N_\varphi = 33 \times 21 \times 20$ 
collocation points.
The bump around the 70th step corresponds to the switch on of the
procedure of convergence towards a given baryon mass.}  
\end{figure}

First of all, we show in Fig.~\ref{f:conv_curve} the convergence
of the iterative procedure described in Sect.~\ref{s:iter}. This convergence
is measured by means of the relative difference $\delta H$ between
two successive steps values of the enthalpy field, as given by
Eq.~(\ref{e:delta_H}). 
The bump around the 70th step corresponds to the
switch on the procedure of convergence towards a given baryon mass,
as described in Sect.~\ref{s:conv_bar_mass}.
One can notice systematic oscillations in the convergence curve
every 8 steps. They result from the fact that the comp-potentials are
updated only every 8 steps, as discussed in Sect.~\ref{s:descrip_step}.
We stop the iterations when the convergence has reached the $\delta H = 10^{-7}$
level (dashed horizontal line in Fig.~\ref{f:conv_curve}).

Next, we show the convergences of the global quantities (i.e.
ADM mass and total angular momentum) for one configuration
when the number of spectral coefficients (or equivalently 
the number of collocation points, cf. Sect.~\ref{s:spectr}) is increased. 
Furthermore, we present the convergence of the relative change in central
energy density along a sequence
when we increase the number of spectral coefficients.
The calculations are performed for the case of $\gamma=2$,
$\kappa=0.0332 \, \rho_{\rm nuc}^{-1} c^2$ 
($\rho_{\rm nuc} :=1.66 \times 10^{17} {\rm kg ~m^{-3}}$);
the baryon mass is $M_B=1.625M_\odot$, which corresponds to 
the compactness $M/R=0.14$ for isolated spherical stars. 
The coordinate separation $d$ is taken to be $60 {\ \rm km}$. 
Six domains are used, with the following parameters
(using the notations of Sect.~\ref{s:domains}): 
$N_{\langle 1 \rangle} = N_{\langle 2 \rangle} = 3$,
$M_{\langle 1 \rangle} = M_{\langle 2 \rangle} = 1$, with 
the same number of coefficients in each domains:
$N_r^{\langle 1 \rangle}(0) = N_r^{\langle 1 \rangle}(1) = \cdots =: N_r$,
$N_\theta^{\langle 1 \rangle}(0) = N_\theta^{\langle 1 \rangle}(1) = 
\cdots =: N_\theta$ and
$N_\varphi^{\langle 1 \rangle}(0) = N_\varphi^{\langle 1 \rangle}(1) 
= \cdots =: N_\varphi$.

\begin{figure}
\centerline{ \epsfig{figure=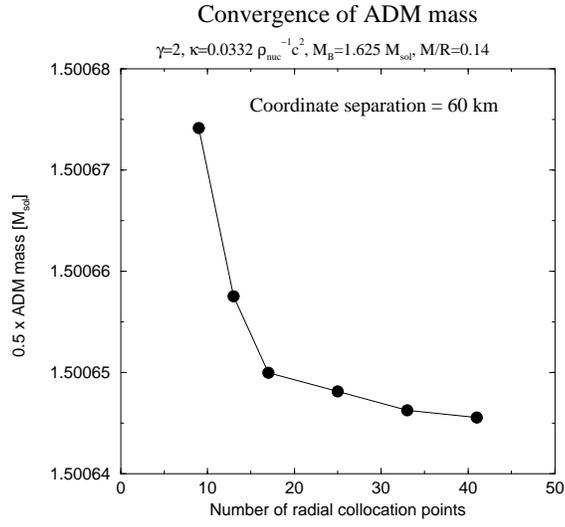,height=7cm} }
\caption[]{\label{f:conv_ADM} 
Convergence of the ADM mass for one of the irrotational relativistic 
models, as the number of collocation points 
(or equivalently of spectral coefficients) is increased.}
\end{figure}

\begin{figure}
\centerline{ \epsfig{figure=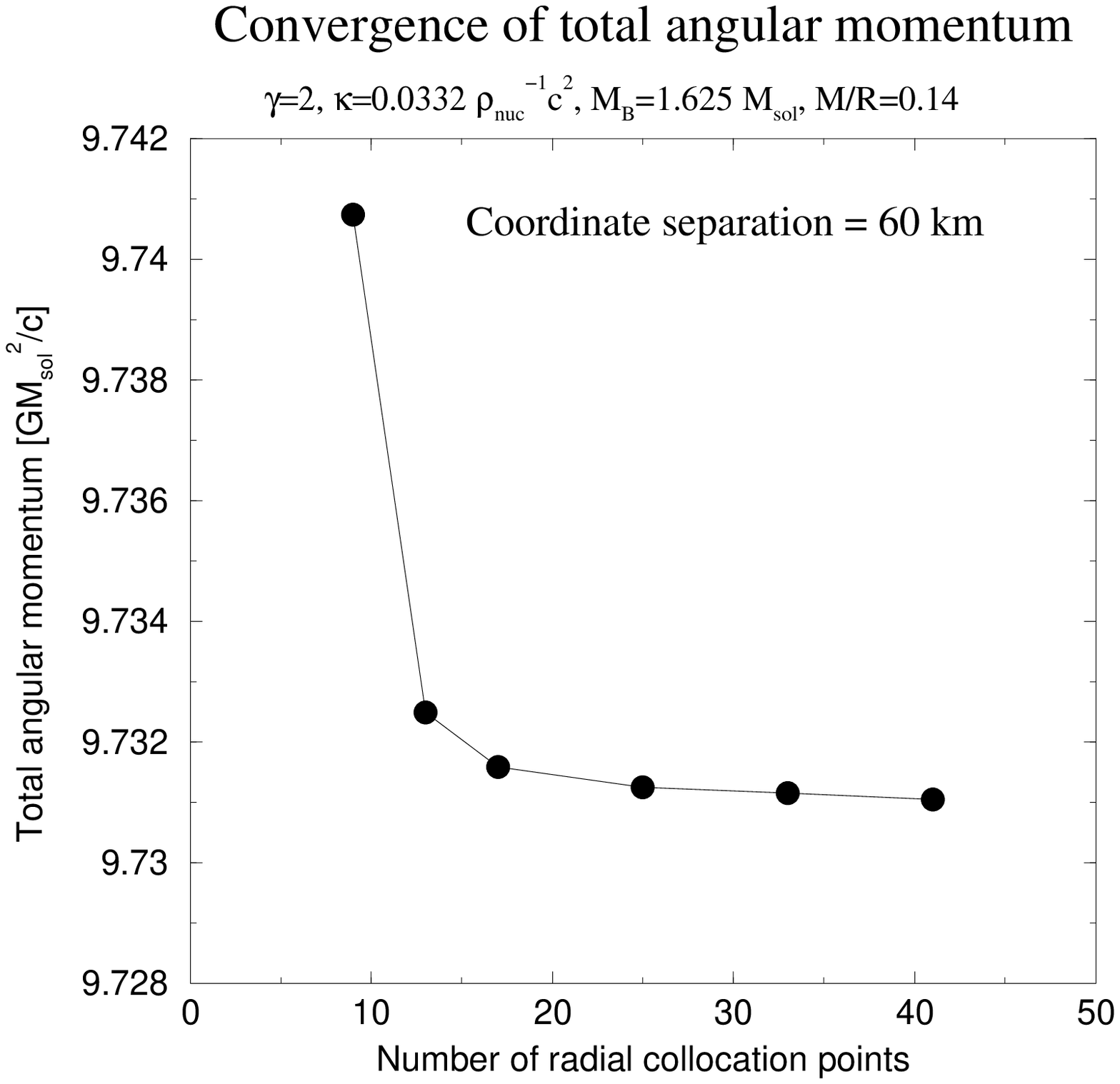,height=7cm} }
\caption[]{\label{f:conv_angmom} 
Same as Fig.~\ref{f:conv_ADM} but for the total angular momentum.}
\end{figure}

\begin{figure}
\centerline{ \epsfig{figure=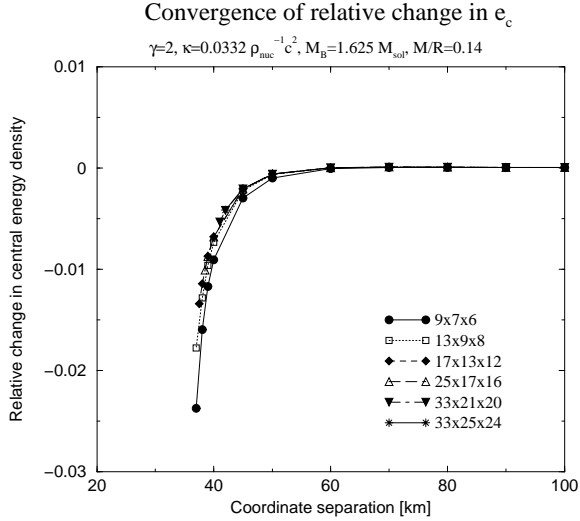,height=7cm} }
\caption[]{\label{f:conv_centdens}
Convergence of the evolution of the central energy density
along a quasiequilibrium sequence, as the number 
$N_r\times N_\theta\times N_\varphi$  of collocation points
(or equivalently of spectral coefficients) is increased.}
\end{figure}

The ADM mass and total angular momentum are shown in
Figs. \ref{f:conv_ADM} and \ref{f:conv_angmom} as functions of $N_r$. 
We used the following numbers of spectral coefficients: 
$N_r \times N_\theta \times N_\varphi = 9 \times 7 \times 6$,
$13\times 9\times 8$, $17\times 13\times 12$, $25\times 17\times 16$,
$33\times 21\times 20$ and $41\times 25\times 24$. 
In Fig. \ref{f:conv_centdens}, we give the relative change
in central energy density along a quasiequilibrium sequence
for various numbers of spectral
coefficients: $(N_r, N_\theta, N_\varphi)=(9, 7, 6)$, $(13, 9, 8)$,
$(17, 13, 12)$, $(25, 17, 16)$, $(33, 21, 20)$, and $(33, 25, 24)$.
We find that these global quantities settle to a constant value
(variations below $\sim 10^{-5}$) for $N_r \geq 25$.

\subsection{Tests in the Newtonian regime}

In order to test Newtonian calculations, 
we compute a  $M_B=10^{-3}M_\odot$ Newtonian sequence
based on a polytropic equation of state with
$\gamma=2$ and $\kappa=0.0332 \, \rho_{\rm nuc}^{-1} c^2$. 
In this case, the central baryon density and the radius of 
infinitely separated stars 
becomes $1.081 \times 10^{-3} \rho_{\rm nuc}$ and $20.57 {\ \rm km}$,
respectively. Note that the Newtonian limit of the polytropic equation
of state (\ref{e:eos_poly_n})-(\ref{e:eos_poly_e}) is obtained for
$H\ll 1$ and reads
\begin{eqnarray}
 n(H) & = &	\left[ {\gamma-1\over \gamma} {m_{\rm B} \over \kappa}\,  H
 	    				\right] ^{1/(\gamma-1)}  \\
 p(H) & = & \kappa \,  n(H)^\gamma  \\
 e(H) & = &  m_{\rm B}  \, n(H) \ . 
\end{eqnarray}

\subsubsection{Virial theorem}

A useful method to estimate the global numerical error in Newtonian
computations is to calculate the relative error in the virial theorem.
This latter states that $2T + W + 3P = 0$, where $T$, $W$, and $P$
denote respectively
the kinetic energy of the binary system, its gravitational potential
energy and the volume integral of the fluid pressure.
We therefore define the virial error as
\be 	\label{eq:virial}
  {\rm Error} = { |2T + W + 3P| \over |W| }  \ . 
\ee

\begin{figure}
\centerline{ \epsfig{figure=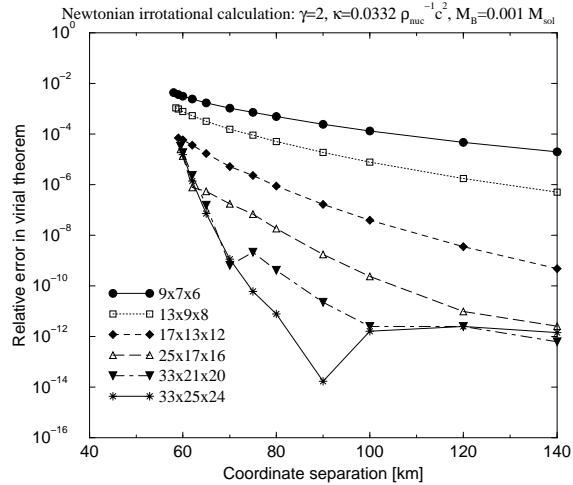,height=7cm} }
\caption[]{\label{f:conv_virial} 
Relative error in the virial theorem
along an evolutionary sequence, for various numbers 
$N_r\times N_\theta\times N_\varphi$  of collocation points
(or equivalently of spectral coefficients).}
\end{figure}

This error estimator is shown along a constant baryon number sequence
in Fig.~\ref{f:conv_virial}.
In order to check the convergence of the numerical method,
we present various cases of increasing number of spectral coefficients:
$N_r \times N_\theta \times N_\varphi = 9\times 7\times 6$,
$13\times 9\times 8$, $17\times 13\times 12$, $25\times 17\times 16$,
$33\times 21\times 20$ and $33\times 25\times 24$.
We used $\delta H=10^{-12}$ as the criterion to end the iteration. 
It is found from Fig.~\ref{f:conv_virial} that for large separations
the relative error converges to $10^{-12}$ when the number of spectral 
coefficients is increased, which is of the order of $\delta H$.
Anyway, one cannot go much further, even if $\delta H$ is lowered 
significantly, because of the use of 15 digits numbers (double precision) 
and the resulting accumulation of round-off errors in 
the arithmetical operations.
Besides, we notice in Fig.~\ref{f:conv_virial} the appearance of 
a rapidly increasing error for closer separations. 
Note however that at the
point of closest approach (cusp point) this error is below
$10^{-4}$ (except for the low numbers of spectral coefficients), which 
is very satisfactory. 
Finally, we see on Fig.~\ref{f:conv_virial} that when we increase the 
number of polar and azimuthal collocation points fixing the number of 
radial ones, the relative error in the virial theorem becomes better around
intermediate separations.

\subsubsection{Comparison with analytic solutions} \label{s:comp_anal}

Until recently the only analytic solutions for binary stars were
constructed with incompressible fluid and belong to the so-called
families of Roche-Riemann or 
Darwin-Riemann ellipsoids\footnote{Note however that these
solutions are not exact for the gravitational potential
of the companion must be truncated to the second order to get
perfectelly ellipsoidal shapes} \cite{Chandra69,Aizen68} (see 
Ref.~\cite{LaiRS93} for a good introduction to these ellipsoidal
solutions). We have
presented elsewhere \cite{BonazGM98a} the comparison with Roche ellipsoids
(the sub-class of Roche-Riemann ellipsoids constituted by synchronized
systems), as a validation of our multi-domain spectral approach
with surface-fitted coordinates. As can be seen from Fig.~6 of 
Ref.~\cite{BonazGM98a} the numerical error is decreasing exponentially 
with the number of spectral coefficients (the so-called {\em evanescent}
error typical of spectral methods), reaching $10^{-9}$ for a Roche
ellipsoid with axis ratios $a_2/a_1 = 0.75$ and $a_3/a_1 = 0.68$. 

The case of compressible fluid bodies has been investigated recently by 
Taniguchi \& Nakamura \cite{TanigN00a,TanigN00b}, who have obtained 
semi-analytic solutions for equilibrium sequences of 
irrotational binary polytropic stars in Newtonian gravity.
For an equal-mass star binary system with $\gamma=2$,
they have produced the following simple equations for the total energy $E$,
the total angular momentum $J$, the orbital angular velocity $\Omega$ 
and the relative change in central baryon density:
\begin{eqnarray}
  E &=&{GM^2 \over R_0} \Bigl[ -1 -{1 \over 2} \Bigl( {R_0 \over d} \Bigr)
  +2 \Bigl( {15 \over \pi^2} -1 \Bigr) \Bigl( {R_0 \over d} \Bigr)^6 \Bigr],
  			\label{e:E_anaTN} \\
  J &=&{1 \over 2} M d^2 \Omega \Bigl[ 1 + {\rm higher~term~than~} O \Bigl(
  \Bigl( {R_0 \over d} \Bigr)^6 \Bigr) \Bigr], 
			\label{e:J_anaTN} \\
  \Omega^2 &=&{2GM \over d^3} \Bigl[ 1 +6 \Bigl( {15 \over \pi^2} -1 \Bigr)
  \Bigl( {R_0 \over d} \Bigr)^5 \Bigr], 
			\label{e:Omega_anaTN} \\
  \delta \rho_c & = & {\rho_c - \rho_{c0} \over \rho_{c0} } =
  -{45 \over 2\pi^2} \Bigl( {R_0 \over d} \Bigr)^6,
			\label{e:drho_anaTN}
\end{eqnarray}
where $d$ is the separation, $\rho_c$ the central density, 
$R_0$ the radius of the spherical star of mass $M$ (i.e. the radius at
infinite separation) and $\rho_{c0}$ the central density of this spherical star. 
These equations are exact up to $O( (R_0/d)^6 )$ and are very valuable
to check the validity of the Newtonian limit of our code, in particular the
solution of equation (\ref{e:psinum2}) for the velocity potential. 
Indeed the Darwin-Riemann solutions could not have been used for testing
this important part of the code
because Eq.~(\ref{e:psinum2}) is degenerate for an 
incompressible fluid ($\zeta = \infty$).

\begin{figure}
\centerline{ \epsfig{figure=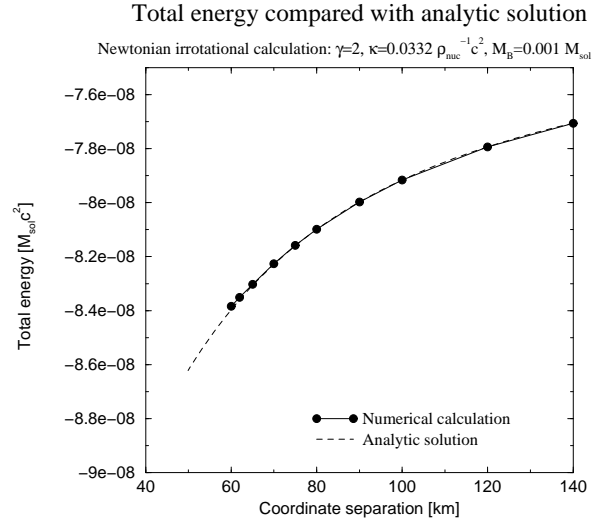,height=7cm} }
\caption[]{\label{f:N_ene} 
Total energy compared with Taniguchi \& Nakamura's analytic solution 
\cite{TanigN00a,TanigN00b} along an evolutionary sequence. Solid and dashed 
lines denote the results of numerical and analytic calculations, respectively.}
\end{figure}

\begin{figure}
\centerline{ \epsfig{figure=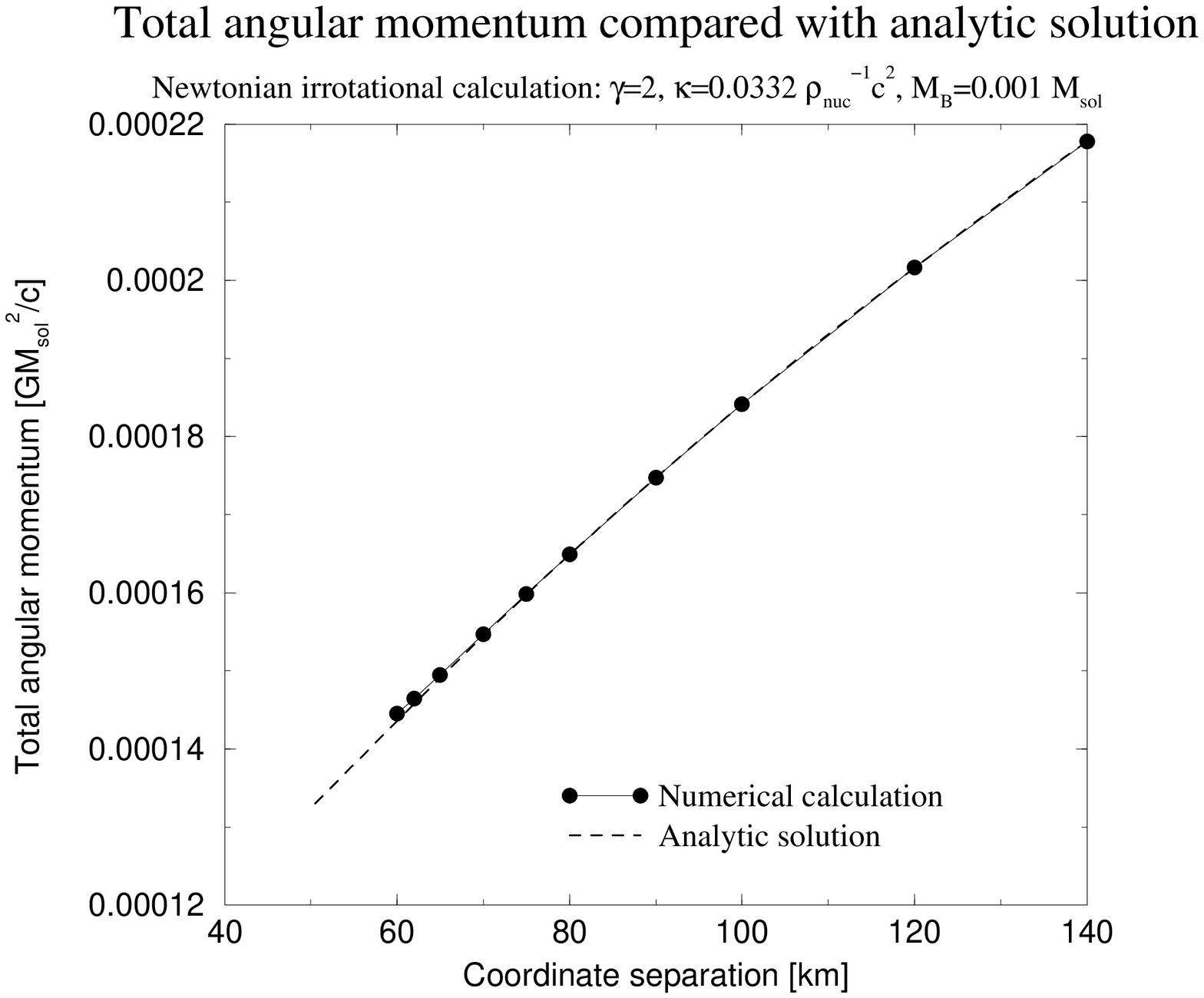,height=7cm} }
\caption[]{\label{f:N_mom} 
Same as Fig.~\ref{f:N_ene} but the total angular momentum.}
\end{figure}

\begin{figure}
\centerline{ \epsfig{figure=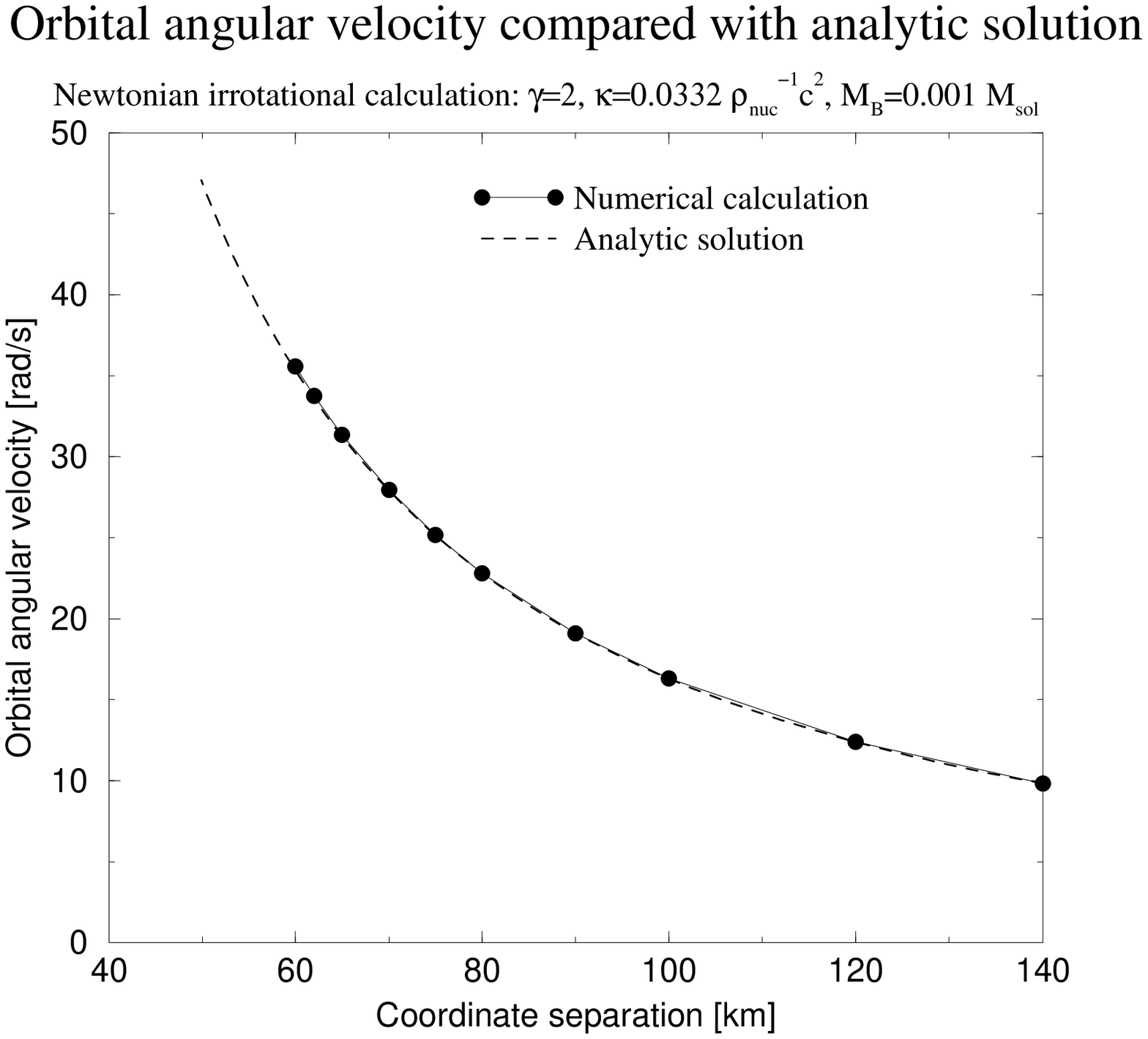,height=7cm} }
\caption[]{\label{f:N_ome} 
Same as Fig.~\ref{f:N_ene} but the orbital angular velocity.}
\end{figure}

\begin{figure}
\centerline{ \epsfig{figure=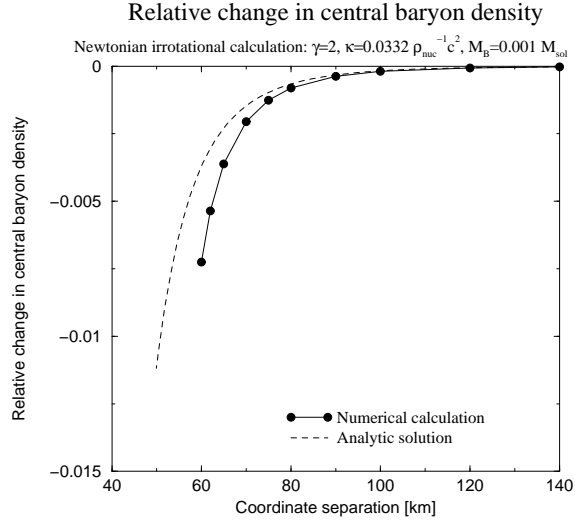,height=7cm} }
\caption[]{\label{f:N_dens} 
Same as Fig.~\ref{f:N_ene} but the relative change in central baryon density.}
\end{figure}

First, we compare our numerical results with Taniguchi \& Nakamura's analytic
solutions for global quantities (such that total energy,
total angular momentum, orbital angular velocity, and
relative change in the central baryon density) 
along an evolutionary sequence in Figs. \ref{f:N_ene} -- \ref{f:N_dens}.
For the numerical computation, we use 
$N_r \times N_\theta \times N_\varphi = 33 \times 25 \times 24$  
spectral coefficients in each domain and
the criterion $\delta H=10^{-12}$ to end the iterations. 
It is found from these figures that the numerical results agree very well
with the analytic ones. Note that the analytic solution ends at the 
contact point, whereas the numerical one ends before (when a cusp appears at the
stellar surface, cf. Sect.~\ref{s:cusp}). 
However the analytic solution, based on an expansion
up to $O( (R_0/d)^6 )$, loses its accuracy for very close separations and 
cannot be used to test the code in this regime.  

\begin{figure}
\centerline{ \epsfig{figure=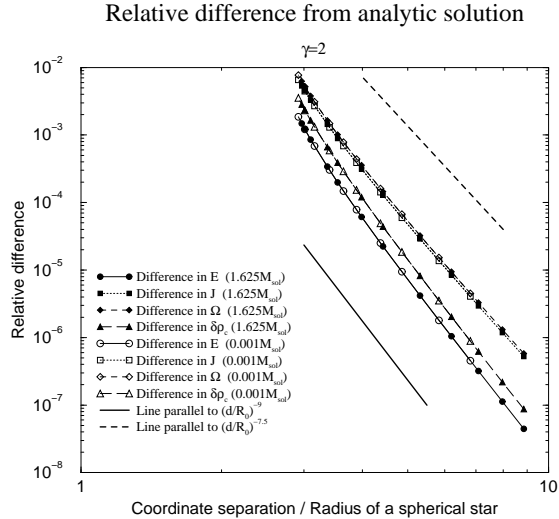,height=7cm} }
\caption[]{\label{f:N_diff} 
Relative differences in total energy $E$, total angular momentum $J$,
orbital angular velocity $\Omega$, and relative change in central
baryon density $\delta\rho_{\rm c}$ when comparing the numerical
solution with Taniguchi \& Nakamura's 
analytic solution \cite{TanigN00a,TanigN00b}
along an equilibrium sequence.
The horizontal axis denotes logarithmically $d/R_0$, where $d$ is the 
separation between the two stellar centers and 
$R_0$ the stellar radius at infinite separation. 
The thick solid and thick dashed lines are reference ones
in order to check the inclinations of the results easily.}
\end{figure}

In order to investigate the discrepancy between the results from the numerical
code and those from Taniguchi \& Nakamura's analytic solution,
we present the relative differences on global
quantities as functions of the separation in a log-log plot in 
Fig.~\ref{f:N_diff}.
The relative differences are defined as follows:
\begin{eqnarray}
  &&{E_{\rm num} - E_{\rm ana} \over GM^2/R_0}, \\
  &&{J_{\rm num} - J_{\rm ana} \over Md^2 \Omega_{\rm Kep}/2}, \\
  &&{\Omega_{\rm num} - \Omega_{\rm ana} \over \Omega_{\rm Kep}}, \\
  &&|\delta \rho_{c:{\rm num}} - \delta \rho_{c:{\rm ana}}|,
\end{eqnarray}
where $\Omega_{\rm Kep}$ is the Keplerian velocity for point mass particles:
\be
  \Omega_{\rm Kep} := \Bigl( {2GM \over d^3} \Bigr)^{1/2}.
\ee
Two reference lines, proportional to $(d/R_0)^{-9}$ and $(d/R_0)^{-7.5}$,
have been drawn in Fig.~\ref{f:N_diff} in order to check the slope
of the results easily.

It is found that for separations closer than $d/R_0=10$, the discrepancies
between numerical and analytic solutions for the energy $E$ and 
the relative change in central density
$\delta \rho_c$ are both proportional to $\sim (d/R_0)^{-9}$,
and become $\sim (d/R_0)^{-13}$ around $d/R_0 \sim 3$.
At first glance, this agreement between the numerical and analytical 
solutions seems too good, because
we know that the next order term missing in Eqs.~(\ref{e:E_anaTN}) and
(\ref{e:drho_anaTN}) is $O(\sim (d/R_0)^{-8})$. 
We interpret the fact that this term does not show up in 
Fig.~\ref{f:N_diff} by the fact that 
it is produced by the octupole deformation, which should be very small. 
Of course, for separations much larger than $d/R_0=10$, the term
proportional to $\sim (d/R_0)^{-8}$ would dominate the inclination
of the lines.

For the residual terms of the angular momentum $J$ and the orbital angular
velocity $\Omega$, we can see that while they are proportional to
$\sim (d/R_0)^{-7}$ around $d/R_0 \sim 10$, the term proportional to
$\sim (d/R_0)^{-8}$
dominates for separations closer than $d/R_0 \sim 6$, and finally goes up to
$\sim (d/R_0)^{-12}$.
We can explain this dependence as follows.
First, the high order expansion of $\Omega$ can be written as 
\be \label{e:Omega_high_order}
  \Omega = \Omega_{\rm Kep} \Bigl[ 1 + O \Bigl( \Bigl( {d \over R_0} \Bigr)^{-5}
  \Bigr) + O \Bigl( \Bigl( {d \over R_0} \Bigr)^{-7} \Bigr)
  + O \Bigl( \Bigl( {d \over R_0} \Bigr)^{-8} \Bigr) +\cdots \Bigr].
\ee
Note here that the second term inside the brackets 
comes from the quadrupole deformation of the star
and is included in the analytic solution [Eq.~(\ref{e:Omega_anaTN})].
After subtracting the analytic solution (\ref{e:Omega_anaTN}) from
expression (\ref{e:Omega_high_order}), there remains the term
$O( (d/R_0)^{-7} )$ as a leading one. Therefore it dominates the behavior
of the curve of the relative difference in $\Omega$ around $d/R_0 \sim 10$.

On the other hand, the angular momentum is expandable as 
\be
  J = {M \over 2} d^2 \Omega \Bigl[ 1 +
  O \Bigl( \Bigl( {d \over R_0} \Bigr)^{-8} \Bigr) +\cdots \Bigr].
\ee
This means that after subtracting the analytic solution (\ref{e:J_anaTN})
and normalizing by $M d^2 \Omega_{\rm Kep} /2$,
the leading term of $J$ comes from $\Omega$,
because this latter has a term proportional to $(d/R_0)^{-7}$.
This explains why the discrepancy curves for $J$ and $\Omega$ have almost the
same behavior.

From the above discussion about the slopes of Fig.~\ref{f:N_diff} curves,
we can conclude that the numerical solution agrees with the semi-analytical
one within the accuracy of this latter, i.e. the increase of the
discrepancy when the separation decreases is due to missing (high order)
terms in the analytic solution (\ref{e:E_anaTN})-(\ref{e:drho_anaTN}). 
Finally, we see from Fig.~\ref{f:N_diff} that even if
the baryon mass is changed by a factor larger than $10^3$, 
the numerical and analytical solutions agree in the very same manner.

\begin{figure}
\centerline{ \epsfig{figure=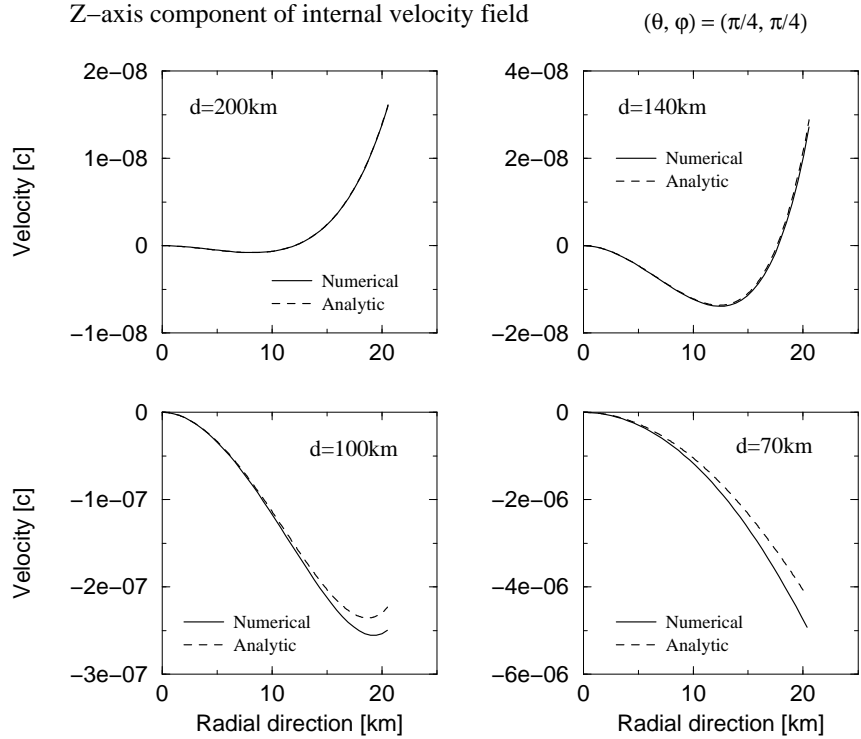,height=10cm} }
\caption[]{\label{f:velo1} 
The $z$-axis component of the internal velocity field in the co-orbiting
frame compared with Taniguchi \& Nakamura's 
analytic solution \cite{TanigN00a,TanigN00b}.
The horizontal line denotes the radial distance from
the center to the surface of star 1, in the direction
$(\theta_{\langle 1 \rangle}, \varphi_{\langle 1 \rangle})=(\pi/4, \pi/4)$. 
The four panels are snapshots at different separations: 200~km, 140~km, 
100~km, and 70~km.
Solid and dashed lines denote the results of numerical and analytic
calculations, respectively.}
\end{figure}

\begin{figure}
\centerline{ \epsfig{figure=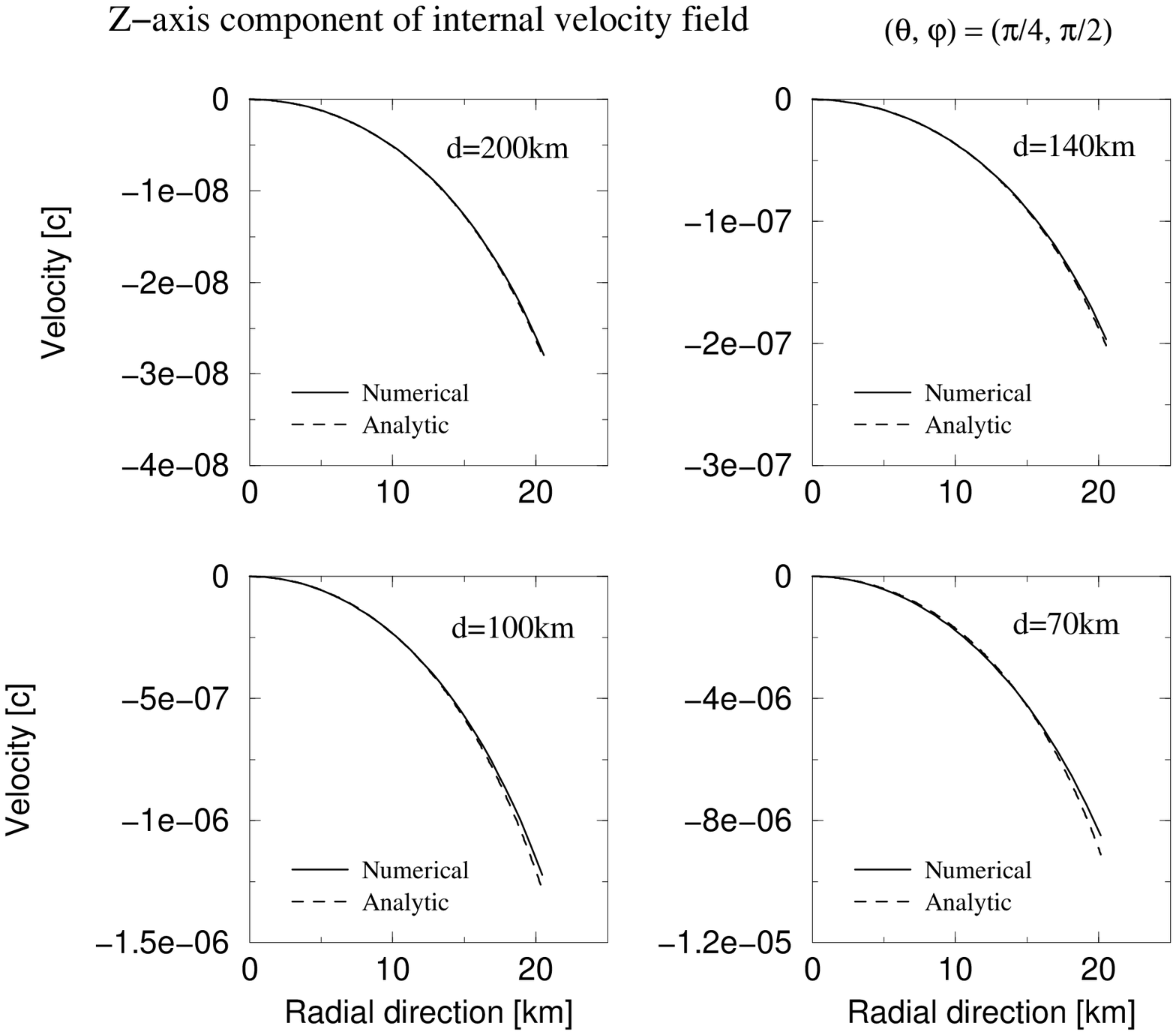,height=10cm} }
\caption[]{\label{f:velo2} 
Same as Fig.~\ref{f:velo1} but for the direction 
$(\theta_{\langle 1 \rangle}, \varphi_{\langle 1 \rangle}) = (\pi/4, \pi/2)$.}
\end{figure}

\begin{figure}
\centerline{ \epsfig{figure=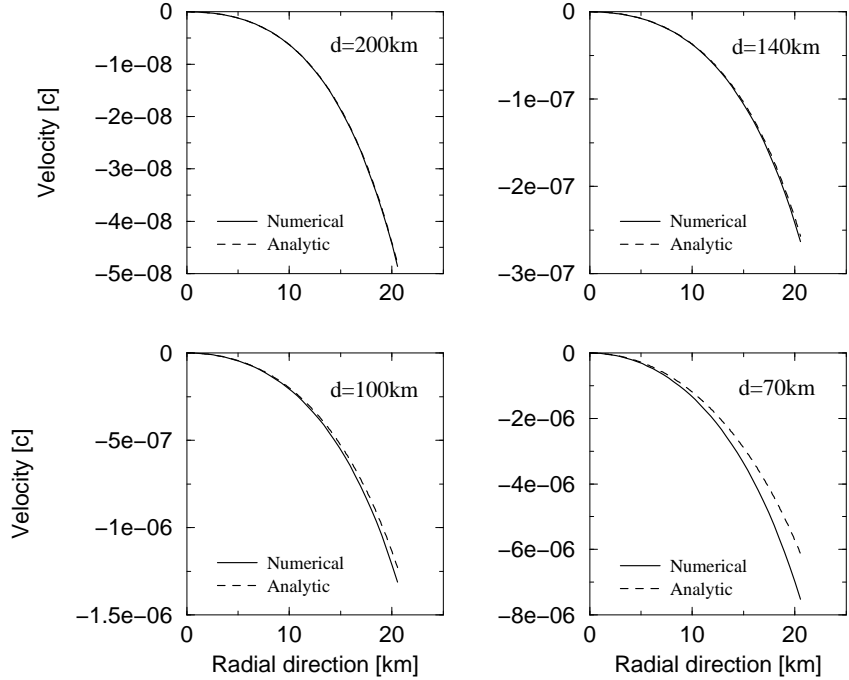,height=10cm} }
\caption[]{\label{f:velo3} 
Same as Fig.~\ref{f:velo1} but for the direction 
$(\theta_{\langle 1 \rangle}, \varphi_{\langle 1 \rangle}) = (\pi/4, 3\pi/4)$.}
\end{figure}

Next, we compare the internal velocity field in the co-orbiting frame
with that of Taniguchi \& Nakamura's analytic solution.
We focus on the velocity component along the orbital axis ($z$-axis),
because it is three orders of magnitude 
smaller than the $x$ and $y$-axis components even for the case of closer
separation, and is therefore a
very valuable quantity  to check whether the equation of continuity 
(\ref{e:psinum2}) is well solved or not. 
In Figs. \ref{f:velo1} -- \ref{f:velo3},
we show the velocity $z$-component as a function
of the radial distance
$r_{\langle 1 \rangle}$ from the center of star 1 along 
three directions:
$(\theta_{\langle 1 \rangle}, \varphi_{\langle 1 \rangle})=$ 
$(\pi/4, \pi/4)$,
$(\pi/4, \pi/2)$, and $(\pi/4, 3\pi/4)$ and for 
the orbital separations $d=200 {\rm \ km}$, 
140~km, 100~km, and 70~km.
It is found that the numerical results agree nicely with those of analytic
calculations. Once again, note that the discrepancy at small separation
comes from the fact that the analytic solution deviates substantially 
from the exact solution.

\subsubsection{Comparison with previous numerical solutions}

As a final test for Newtonian computations, we compare our results with
those of Uryu \& Eriguchi \cite{UryuE98b} for polytropic equation
of state with $\gamma = 5/3$, $2$ and $3$, corresponding to 
polytropic indices $n=1.5$, $1$ and $0.5$ respectively ($\gamma = 1 + 1/n$).  
The comparison is presented in Table~\ref{table1}, where the
upper lines for each configuration are the results of Uryu \& Eriguchi
and the lower ones are ours.
We have chosen the configurations $\tilde{d}=3.6$ in Tables~2, 4, and 
5 of Uryu \& Eriguchi \cite{UryuE98b}. 
Here, our results are calculated by using 
$N_r \times N_\theta \times N_\varphi = 33 \times 25 \times 24$  
spectral coefficients in each of the 6 domains
and we adopted the same definitions as in 
Uryu \& Eriguchi's article \cite{UryuE98b}, namely
\begin{eqnarray}
  \bar{d}_G &:=&{d_G \over R_0}, \\
  \bar{\Omega} &:=&{\Omega \over (\pi G \bar{\rho}_0)^{1/2}}, \\
  \bar{J} &:=&{J \over (GM^3 R_0)^{1/2}}, \\
  \bar{E} &:=&{E \over GM^2/R_0} \ , 
\end{eqnarray}
where $d_G$ is the distance between the two stellar centers of mass
and $\bar{\rho}_0 := M / (4\pi R_0^3 /3)$.   

One can see from Table~\ref{table1} that our results coincide with those of
Uryu \& Eriguchi within 0.3 \% for physical values such as
the total energy, the total angular momentum and the orbital angular
velocity. Note that the label $\tilde{d}$ of Uryu \& Eriguchi \cite{UryuE98b}
configurations is the orbital separation between
the geometrical centers of two stars normalized by the geometrical radius
of the star along the $x$-axis.
In our computation, since the geometrical separation is obtained after
calculation, we cannot fix $\tilde{d}$ initially. 
Therefore we use the corresponding
separation between the centers of mass of two stars which Uryu \& Eriguchi
gave in their paper\cite{UryuE98b} as the orbital separation
between the {\it centers} of two stars $\bar{d}=d/R_0$.
Although our definition of the center of the star,
which is the location of the maximum enthalpy (Sect.~\ref{s:domains}),
is different from the center of mass, the relative difference between
these centers is only about 0.01 \% around $\tilde{d}=3.6$.

\begin{table}
\caption{\protect Comparison with the results of Uryu \& Eriguchi (1998).}
 \begin{center}
  \begin{tabular}{ccccc}
  Separation&$\bar{\Omega}$&$\bar{J}$&$\bar{E}$ \\ \hline
  \multicolumn{5}{c}{$\gamma=3$  ($n=0.5$)} \\ \hline
  $\bar{d}_G=3.804$&0.2219&1.385&-1.241 \\
  $\bar{d}=3.804$&0.2211&1.385&-1.242 \\ \hline
  \multicolumn{5}{c}{$\gamma=2$  ($n=1$)} \\ \hline
  $\bar{d}_G=3.753$&0.2259&1.371&-1.133 \\
  $\bar{d}=3.753$&0.2252&1.373&-1.133 \\ \hline
  \multicolumn{5}{c}{$\gamma=5/3$ ($n=1.5$)} \\ \hline
  $\bar{d}_G=3.726$&0.2279&1.364&-0.9921 \\
  $\bar{d}=3.726$&0.2274&1.367&-0.9911 \\
  \end{tabular}
 \end{center}
 \label{table1}
\end{table}%

\subsection{Test of the Newtonian limit of relativistic calculations}

\begin{figure}
\centerline{ \epsfig{figure=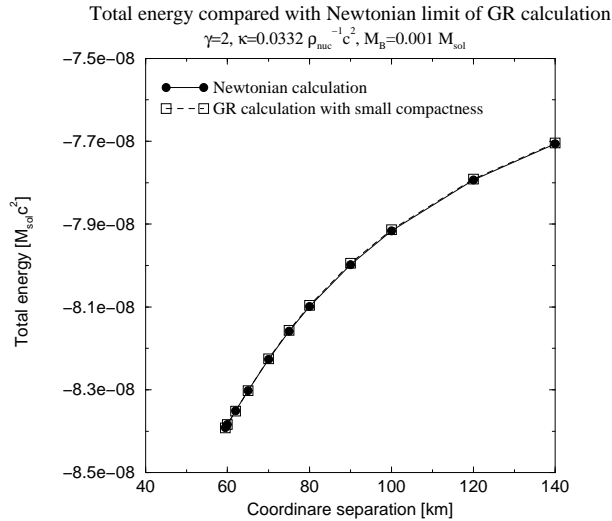,height=7cm} }
\caption[]{\label{f:limGR_ene} 
Total energy of a relativistic sequence of small compactness
($M/R=7.18\times10^{-5}$) compared with that of that of a Newtonian 
sequence of the same mass.
Solid line with filled circles denotes the Newtonian computation and
dashed line with squares denotes the relativistic one.}
\end{figure}

\begin{figure}
\centerline{ \epsfig{figure=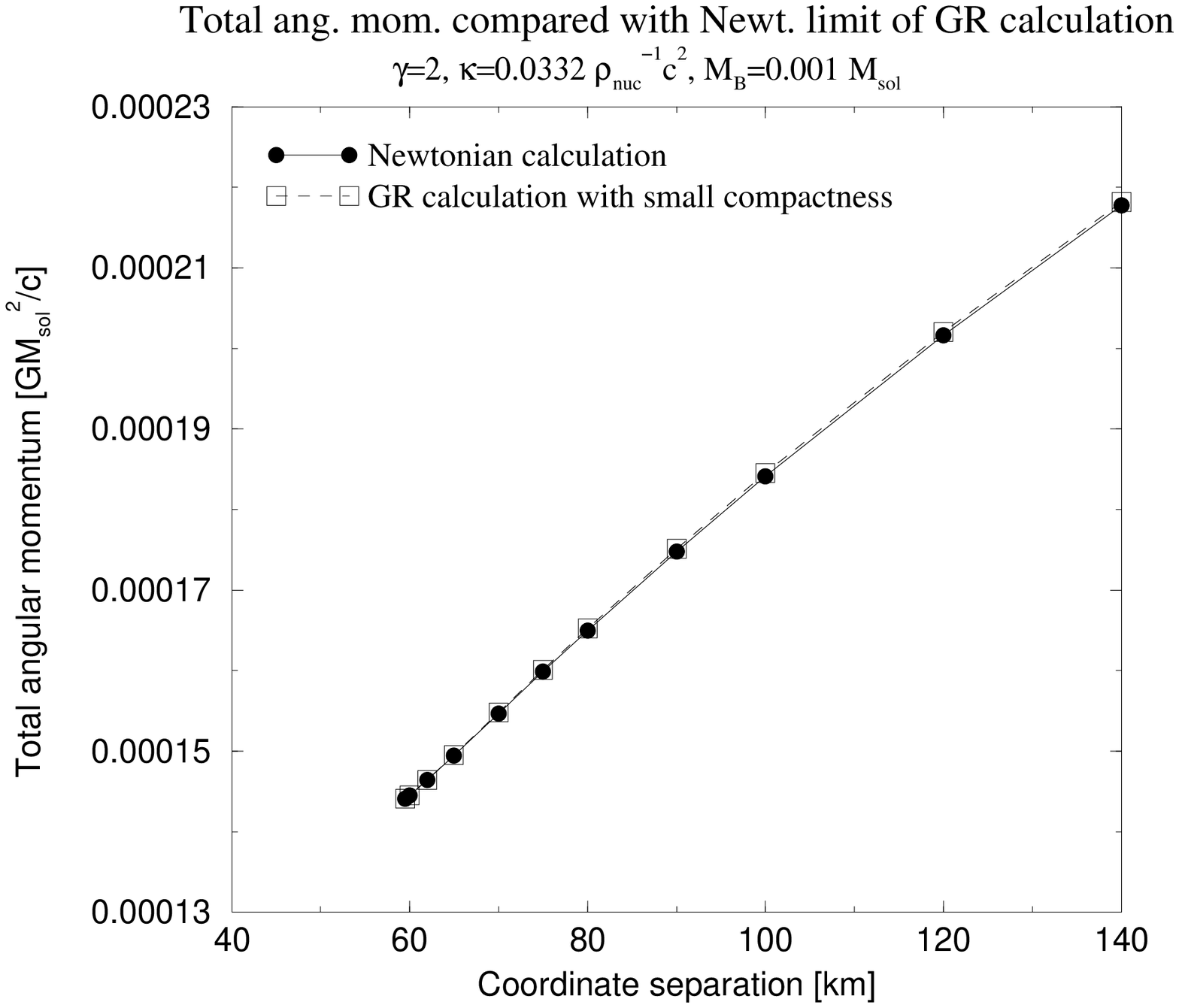,height=7cm} }
\caption[]{\label{f:limGR_mom} 
Same as Fig.~\ref{f:limGR_ene} but for the total angular momentum.}
\end{figure}

\begin{figure}
\centerline{ \epsfig{figure=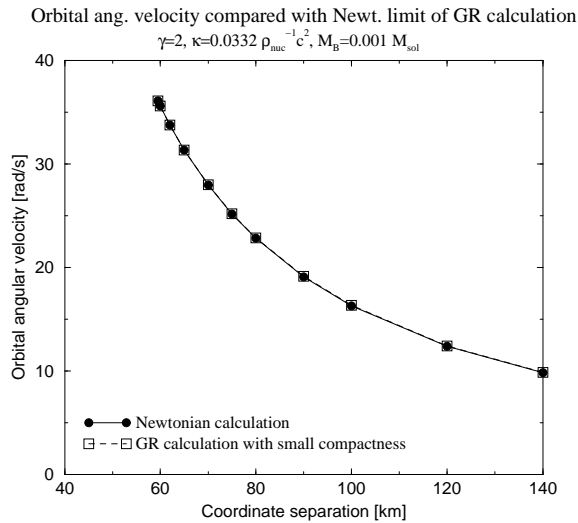,height=7cm} }
\caption[]{\label{f:limGR_ome} 
Same as Fig.~\ref{f:limGR_ene} but for the orbital angular velocity.}
\end{figure}

We have made many tests of the code in the Newtonian regime up to now, 
so that we are rather confident in the accuracy of the Newtonian part
of the code. 
As a next step, we compare the results of relativistic calculations
with small compactness ($M/R=7.18\times10^{-5}$)
with those of Newtonian ones.
In Figs. \ref{f:limGR_ene} -- \ref{f:limGR_ome},
the total energy, the total angular momentum and the orbital angular
velocity are shown along a sequence. We use
$N_r \times N_\theta \times N_\varphi = 25 \times 17 \times 16$  
spectral coefficients in each domain. 
It appears clearly that the results of the small compactness relativistic
computation coincide with those of the Newtonian computation, as it should be.

\subsection{Comparison with previous relativistic numerical solutions}

\subsubsection{Corotating case}

As a check of for relativistic computations, we compare our results for
corotating configurations
with those of Baumgarte et al.\cite{BaumgCSST98}.
We have chosen Tables III and VI of their paper, and
compare the results of two different separations $z_A=0.2$ and 0.3
in each table.
We use $N_r \times N_\theta \times N_\varphi = 33 \times 25 \times 24$  
spectral coefficients in each domain
and the criterion $\delta H=10^{-7}$ to end the computation of one
configuration. 
We adopt the same equation of state (polytropic with
$\gamma=2$), the same value of the separation $r_C=d/2$ and the same value
of the baryon mass $\bar{M}_0$.
These results are shown in Table~\ref{table2} where
the upper lines for each configuration denote
the results of  Baumgarte et al.,
and the lower ones correspond to our results.
We find a relative discrepancy of 2 \%
on $\bar{\Omega}$, 4.5 \% on $q^{\rm max}$, 0.07 \% on $\bar{M}$,
0.6 \% on $\bar{J}$, 4.5 \% on $\bar{r}_A$, and 1.5 \% on $\bar{r}_B$.

\begin{table}
\caption{Comparison with the results of Baumgarte et al. (1998)}
 \begin{center}
  \begin{tabular}{ccccccccc}
  $z_A$&$\bar{M}_0$&$q^{\rm max}$&$\bar{M}$&$\bar{J}$&$\bar{\Omega}$&
  $\bar{r}_A$&$\bar{r}_C$&$\bar{r}_B$ \\ \hline
  \multicolumn{9}{c}{TABLE III $(M/R=0.05)$} \\ \hline
  0.20&0.0595&0.0284&0.057815&0.01109&0.048&0.591&1.791&2.959 \\
      &      &0.0280&0.057816&0.01113&0.048&0.582&  &2.923 \\ \hline
  0.30&      &0.0288&0.057836&0.01155&0.037&0.975&2.118&3.251 \\
      &      &0.0285&0.057836&0.01161&0.038&0.968&  &3.217 \\ \hline
  \multicolumn{9}{c}{TABLE VI  $(M/R=0.15)$} \\ \hline
  0.20&0.1534&0.1303&0.140859&0.04174&0.116&0.413&1.244&2.067 \\
      &      &0.1242&0.140774&0.04194&0.117&0.395&   &2.037 \\ \hline
  0.30&      &0.1341&0.140971&0.04268&0.092&0.682&1.477&2.273 \\
      &      &0.1286&0.140874&0.04294&0.093&0.668&   &2.249 \\
  \end{tabular}
 \end{center}
 \label{table2}
\end{table}%

\subsubsection{Irrotational case}

For irrotational relativistic configurations, 
a detailed comparison with Uryu \& Eriguchi results 
\cite{UryuE00,UryuSE00} is underway
\cite{GourgU00}. For the purpose of the present article, we have compared
only the cusp point configuration of a $M/R=0.14$ $\gamma=2$ polytropic
sequence as given in the last but one line of Table~IV of 
Ref.~\cite{UryuE00} and
the last line of our Table~\ref{table4} below. The agreement is
very satisfactory: the relative discrepancy is below $0.1 \%$ on $M$, 
$0.4\%$ on $\Omega M_{\rm B}$, $2.1\%$ on $J/M^2$ and $2.1\%$ on the
semi-diameter $a_0$ of the stars [Eq.~\ref{e:def_a_0}) below].

\section{Results for $\gamma=2$ polytropes} \label{s:results}

The tests of the code being successfully passed, we employ the code to
compute an irrotational relativistic sequence based on the polytropic 
equation of state (\ref{e:eos_poly_n})-(\ref{e:eos_poly_e})
with $\gamma=2$. 
We are using $\kappa=0.0332 \rho_{\rm nuc}^{-1} c^2$, and consider
the compactness parameter $M/R=0.14$ at infinite separation. 
This results in the baryon mass $M_{\rm B}=1.625 M_\odot$.

The computational parameters are as follows:
six domains are used, such that 
(using the notations of Sect.~\ref{s:domains}): 
$N_{\langle 1 \rangle} = N_{\langle 2 \rangle} = 3$,
$M_{\langle 1 \rangle} = M_{\langle 2 \rangle} = 1$, with 
the same number of coefficients in each domain:
$N_r \times N_\theta \times N_\varphi = 33 \times 21 \times 20$.  
The criterion to end the computation of one configuration is set
to $\delta H=10^{-7}$. 
A special treatment has been performed for the closest configuration,
because of the existence of a cusp on the stellar surface (Sect.~\ref{s:cusp}):  
$N_r \times N_\theta \times N_\varphi = 25 \times 17 \times 16$ 
coefficients have been used along with the enthalpy 
gradient threshold $\chi_{\rm fr} = 0.55$, resulting in a frozen mapping. 

\begin{figure}
\centerline{ \epsfig{figure=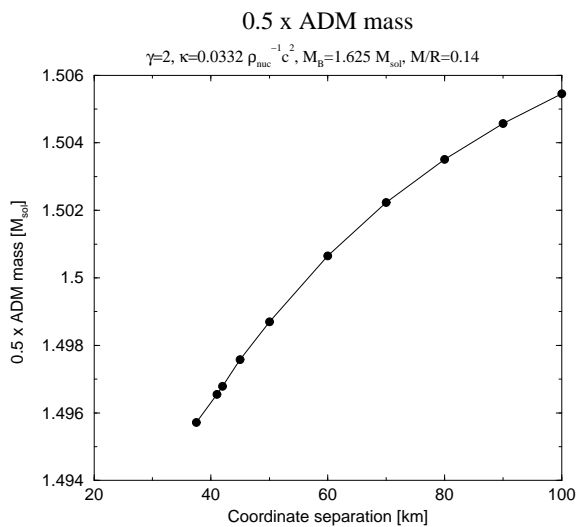,height=7cm} }
\caption[]{\label{f:GR_ADM} 
Half of the ADM mass of the binary system as a function of the coordinate
separation for an evolutionary sequence of relativistic irrotational stars. 
}
\end{figure}

\begin{figure}
\centerline{ \epsfig{figure=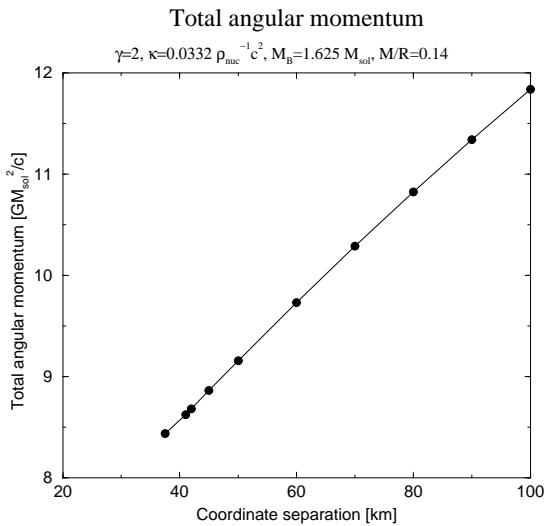,height=7cm} }
\caption[]{\label{f:GR_mom} 
Same as Fig.~\ref{f:GR_ADM} but for the total angular momentum.
}
\end{figure}

In Figs. \ref{f:GR_ADM} and \ref{f:GR_mom}, the half of the ADM mass
and the total angular momentum of the binary system, as defined in
Sect.~\ref{s:global}, are shown along the evolutionary sequence.
This sequence ends at around $d=37.5~ {\rm km}~(f=380 {\ \rm Hz})$
where a cusp appears on the surface of the stars. 

One can see from these figures that there is no turning point
for the $\gamma=2$ case. This result agrees with that of
Uryu \& Eriguchi\cite{UryuE00}.

An important result of this computation has already been 
presented in Ref.~\cite{BonazGM99a}, namely
the central energy density remains rather constant (with a slight increase
below 0.01 \%) and finally decreases (see Fig. \ref{f:conv_centdens}).
As discussed in the Introduction, this result makes the collapse of the
individual neutron stars to black hole very unlikely prior to the merger.  

We summarize the results about the sequence in Table \ref{table3},
listing the ADM mass, the total angular momentum,
the orbital angular velocity, the axis ratios,
and the relative change in central energy density
along the quasiequilibrium sequence.
Since this is the first table presented for a sequence of 
relativistic irrotational binary neutron stars, 
we give a rather large number of digits
in order to compare with the results of other works from now on.
Note that we are using the following values of the fundamental constants:
$G = 6.6726\times 10^{-11} {\ \rm m}^3 {\rm kg}^{-1} {\rm s}^{-2}$,
$c=2.99792458\times 10^8 {\ \rm m\, s}^{-1}$ and 
$M_\odot=1.989\times 10^{30} {\ \rm kg}$.

\begin{table}
\caption{Half of ADM mass $M$, total angular momentum $J$,
orbital angular velocity $\Omega$, axis ratios,
and relative change in central energy density
along a $M_{\rm B}=1.625\, M_\odot$ quasiequilibrium sequence
constructed upon a $\gamma=2$ polytropic EOS. 
$a_1$, $a_2$, and $a_3$ denote the coordinate lengths
parallel to the semi-major axes $x$, $y$, and $z$, respectively.
$a_{1,{\rm opp}}$ is the length in the direction opposite to the companion star.
}
 \begin{center}
  \begin{tabular}{rcccccccr}
  $d \ [{\rm km}]$&$0.5 \times M \ [M_\odot]$&$J \ [GM_\odot^2/c]$&
  $\Omega \ [{\rm rad/s}]$&$\Omega/(2\pi) \ [{\rm Hz}]$&
  $a_2/a_1$&$a_3/a_1$&$a_{1,{\rm opp}}/a_1$&$(e_c - e_{c,\infty})/e_{c,\infty}$
  \\ \hline
  100&1.50545&11.8370&597.24&95.054&0.99100&0.99367&0.99319&4.0606e-05 \\
  90 &1.50457&11.3403&695.15&110.64&0.98839&0.99139&0.99234&6.0695e-05 \\
  80 &1.50351&10.8243&823.17&131.01&0.98445&0.98788&0.99106&6.9369e-05 \\
  70 &1.50223&10.2880&996.14&158.54&0.97811&0.98210&0.98903&8.4666e-05 \\
  60 &1.50065&9.73115&1239.9&197.34&0.96687&0.97171&0.98466&3.2735e-05 \\
  50 &1.49870&9.15576&1603.5&255.20&0.94402&0.95037&0.97369&-5.9816e-04 \\
  45 &1.49758&8.86296&1858.8&295.84&0.92226&0.92999&0.96263&-2.0684e-03 \\
  42 &1.49679&8.68172&2041.4&324.89&0.90315&0.91100&0.95408&-4.2246e-03 \\
  41 &1.49655&8.62425&2111.0&335.98&0.89206&0.89940&0.95076&-5.4239e-03 \\
 37.5&1.49572&8.43623&2389.7&380.33&0.81445&0.82752&0.91949&-1.2238e-02 \\
\end{tabular}
 \end{center}
 \label{table3}
\end{table}%

For comparison purposes with other works, we also 
give Table~\ref{table4} in which
the physical quantities are normalized by using the equation of state
constants $\kappa$ and $\gamma$
to set a length scale $R_{\rm poly}$ according to
\be
	R_{\rm poly} := \kappa^{1\over 2(\gamma-1)} \ . 
\ee
We can therefore define the same dimensionless quantities as 
Baumgarte et al. \cite{BaumgCSST98} and Uryu \& Eriguchi \cite{UryuE00}:
$ \bar{M}_{\rm B} = M_{\rm B} / R_{\rm poly}$, 
$  \bar{M} = M / R_{\rm poly}$, 
$ \bar{J} = J / R_{\rm poly}^2$,
$\bar{\Omega} =  R_{\rm poly} \Omega$, 
$\bar{d} = d / R_{\rm poly}$,
$\bar{d}_G =  d_G / R_{\rm poly}$, 
$\bar{a}_0 = a_0 / R_{\rm poly}$, 
where $d_G$ is the distance between the ``centers of mass'' of each stars
as defined by Eq.~(107) of Uryu \& Eriguchi \cite{UryuE00}:
\be
  d_G := \left| {1 \over M_{\rm B}^{\langle 1 \rangle}}
		\int_{\rm star\,  1} A^3 \Gamma_{\rm n} n X  d^3 x
	- {1 \over M_{\rm B}^{\langle 2 \rangle}}
		\int_{\rm star\,  2} A^3 \Gamma_{\rm n} n X  d^3 x	
	\right|
\ee
and $a_0$ is half of the coordinate length of a star along the $X$ axis:
\be \label{e:def_a_0}
  a_0 = {1\over 2} \, \left| X_{\rm max} - X_{\rm min} \right|.
\ee
This latter quantity is denoted $R_0$ by Uryu \& Eriguchi \cite{UryuE00}.

\begin{table}
\caption{Dimensionless ADM mass $\bar{M}$, total angular momentum $\bar{J}$,
orbital angular velocity $\bar{\Omega}$, and
half of the coordinate length along the $X$ axis $\bar{a}_0$
along the $\bar{M}_{\rm B}=0.146202$ quasiequilibrium sequence 
presented in Table~\ref{table3}. 
}
 \begin{center}
  \begin{tabular}{cccccccc}
  $\bar{d}$& $\bar{d}_G$& $0.5 \times \bar{M}$& $\bar{J}$& $J/M^2$& 
  $\bar{\Omega}$& $\Omega\, M_{\rm B}$&  $\bar{a}_0$ \\ \hline
  6.0927& 6.0924& 0.135446& 9.58174e-02& 1.30573& 3.2698e-02& 4.7805e-03& 0.81089 \\
  5.4835& 5.4830& 0.135367& 9.17965e-02& 1.25239& 3.8058e-02& 5.5642e-03& 0.80934 \\
  4.8742& 4.8736& 0.135272& 8.76195e-02& 1.19708& 4.5067e-02& 6.5889e-03& 0.80770 \\
  4.2649& 4.2642& 0.135156& 8.32782e-02& 1.13973& 5.4536e-02& 7.9733e-03& 0.80623 \\
  3.6556& 3.6546& 0.135014& 7.87708e-02& 1.08031& 6.7883e-02& 9.9246e-03& 0.80542 \\
  3.0464& 3.0447& 0.134839& 7.41131e-02& 1.01907& 8.7788e-02& 1.2835e-02& 0.80796 \\
  2.7417& 2.7396& 0.134738& 7.17429e-02& 0.98796& 1.0177e-01& 1.4879e-02& 0.81489 \\
  2.5589& 2.5565& 0.134667& 7.02758e-02& 0.96878& 1.1176e-01& 1.6339e-02& 0.82583 \\
  2.4980& 2.4954& 0.134645& 6.98107e-02& 0.96268& 1.1557e-01& 1.6897e-02& 0.83271 \\
  2.2848& 2.2810& 0.134571& 6.82887e-02& 0.94273& 1.3083e-01& 1.9127e-02& 0.88016 \\
  \end{tabular}
 \end{center}
 \label{table4}
\end{table}

\begin{figure}
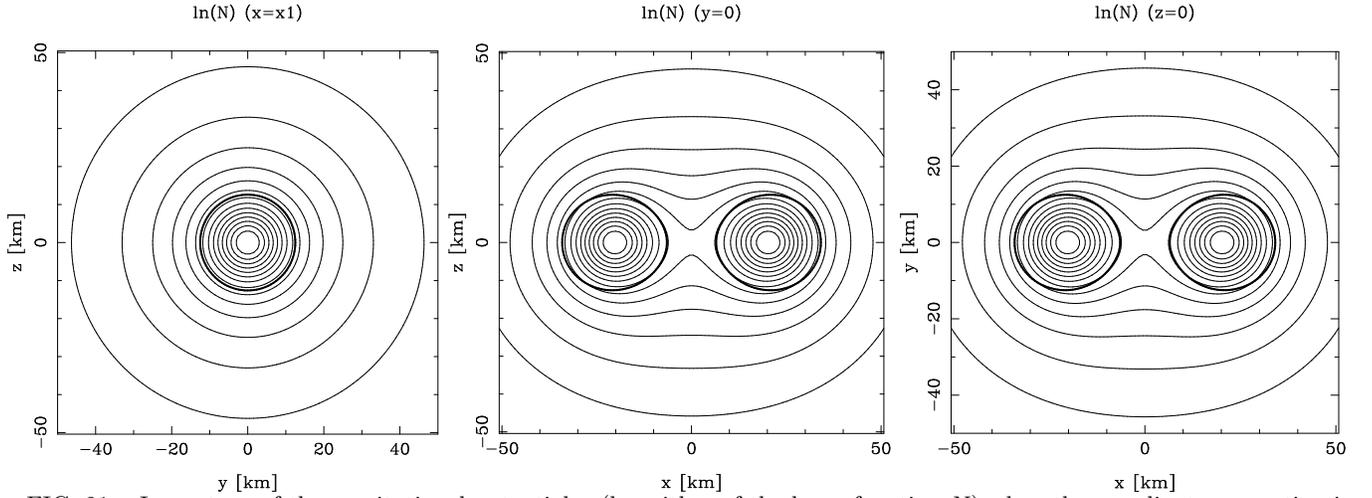

\centerline{ \epsfig{figure=fig_lapse_x.eps,height=6.5cm}
             \epsfig{figure=fig_lapse_y.eps,height=6.5cm}
             \epsfig{figure=fig_lapse_z.eps,height=6.5cm}  }
\caption[]{\label{f:lapse}
Isocontour of the gravitational potential $\nu$ (logarithm of the 
lapse function $N$) when the coordinate separation is 41~km.
The plots are cross sections of the $X=-20.5~{\ \rm km}$, $Y=0$, and $Z=0$
planes. The thick solid lines denote the surfaces of the stars.
}
\end{figure}

Finally, let us show some figures about metric,
density, and internal velocity fields.
The lapse function $N$ is represented in Fig. \ref{f:lapse}.
The coordinate system is the $(X, Y, Z)$ one defined in Sect.~\ref{s:domains}
and the coordinate separation is 41~km (last but one line in Tables~\ref{table3}
and \ref{table4}). 
At this separation, the central value of $N$ of each star is 0.6416.

\begin{figure}
\centerline{ \epsfig{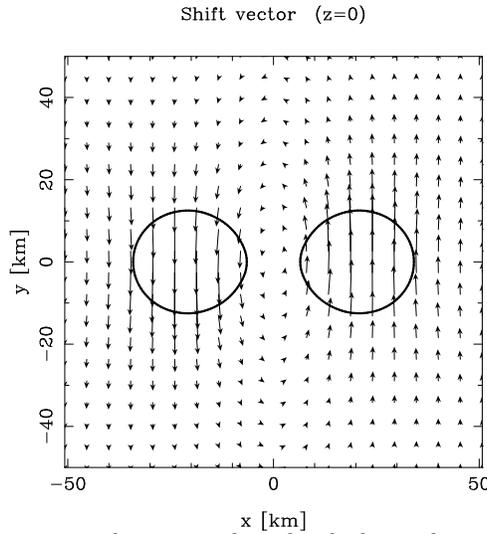} }
\caption[]{\label{f:shift} 
Shift vector $\mib{N}$ of non-rotating coordinates in the orbital plane
when the coordinate separation is 41~km.
The thick solid lines denote the surfaces of the stars.
}
\end{figure}

The shift vector $\mib{N}$ 
of non-rotating coordinates [defined by Eq.~(\ref{e:shift_non_rot})] is shown 
in Fig. \ref{f:shift}.
The plot is a cross section of the orbital plane.

\begin{figure}
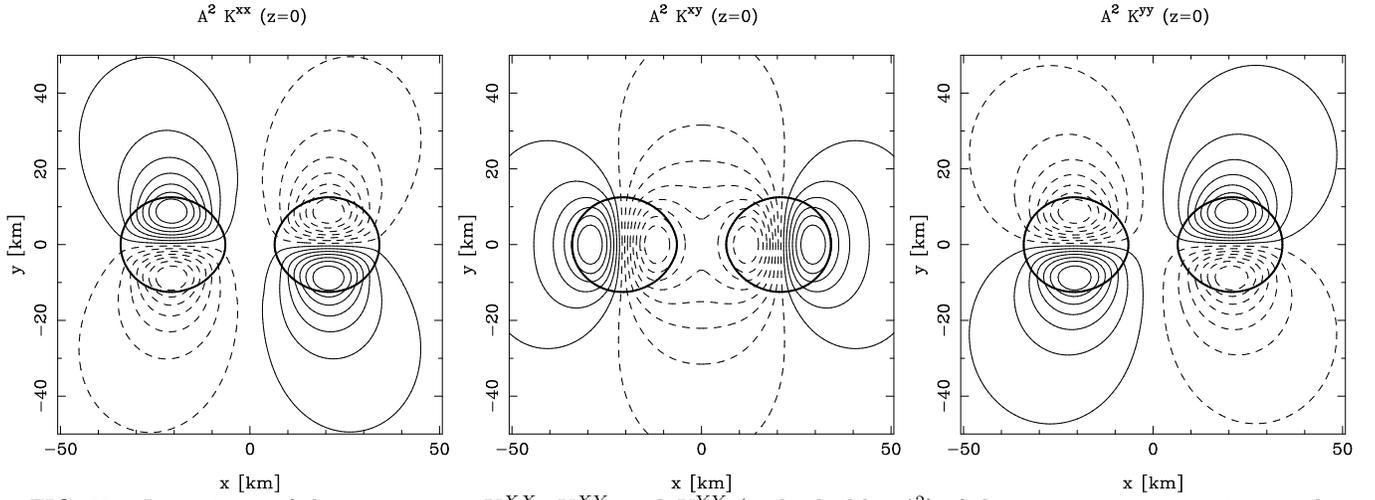

\centerline{ \epsfig{figure=fig_extrinsic_xx.eps,height=6.5cm}
             \epsfig{figure=fig_extrinsic_xy.eps,height=6.5cm}
             \epsfig{figure=fig_extrinsic_yy.eps,height=6.5cm}  }
\caption[]{\label{f:extrinsic} 
Isocontours of the components $K^{XX}$, $K^{XY}$, and $K^{YY}$
(multiplied by $A^2$) 
of the extrinsic curvature tensor when the  coordinate separation is 41~km.
The plots are cross sections of the orbital plane ($Z=0$).
Solid (resp. dashed) lines denotes positive (resp. negative) values. 
The thick solid lines mark the surfaces of the stars.
}
\end{figure}

The $K^{XX}$, $K^{XY}$, and $K^{YY}$ components of the extrinsic curvature
tensor of the hypersurfaces $t={\rm const}$ are shown
in Fig. \ref{f:extrinsic}.
The values in the figures are multiplied by the square conformal factor
$A^2$, and the plots are cross section of the orbital plane.

\begin{figure}
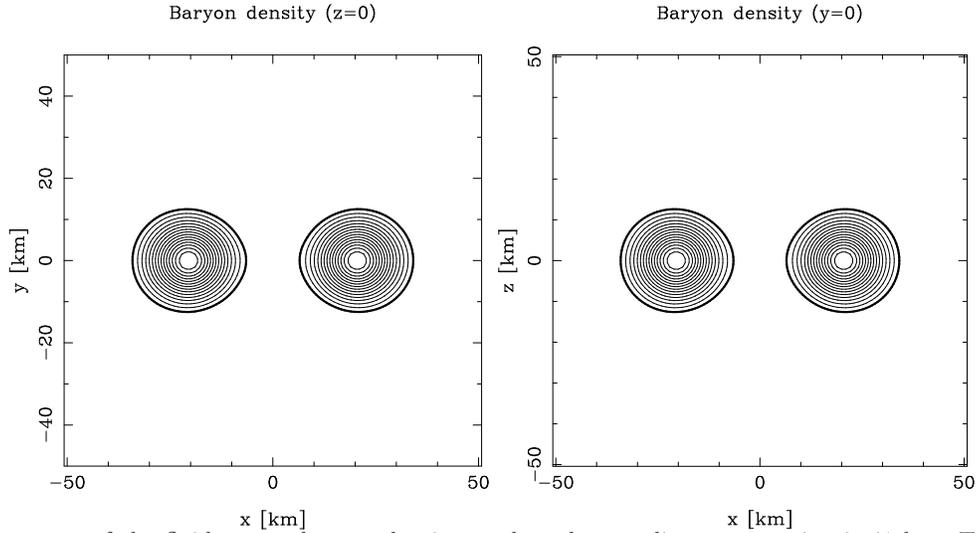

\centerline{ \epsfig{figure=fig_baryon_xy.eps,height=7cm}
             \epsfig{figure=fig_baryon_xz.eps,height=7cm}  }
\caption[]{\label{f:baryon_density} 
Isocontours of the fluid proper baryon density $n$ when
the coordinate separation is 41~km.
The plots are cross sections of $Z=0$ and $Y=0$ planes.
The thick solid lines denote the surfaces of the stars.
}
\end{figure}

The baryon density field in the fluid frame
is shown in Fig.~\ref{f:baryon_density}.
The plots are cross sections of $Z=0$ and $Y=0$ planes.

\begin{figure}
\centerline{ \epsfig{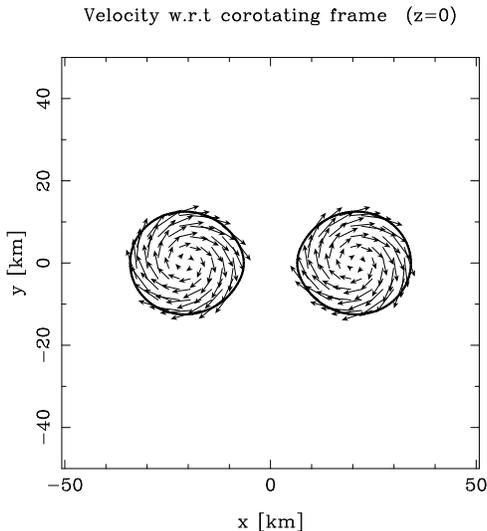} }
\caption[]{\label{f:internal_velo} 
Internal velocity field with respect to the co-orbiting frame
in the orbital plane when the coordinate separation is 41~km.
The thick solid lines denote the surfaces of the stars.
}
\end{figure}

Finally, we show in Fig.~\ref{f:internal_velo}
the internal velocity field in the co-orbiting frame,
or more precisely the orthogonal projection in the
$\Sigma_t$ hypersurface of the vector field $\mib{V}$ given by
Eq.~(\ref{e:3-vitV}). Note that this vector field is tangent to
the surface of the stars, as it should be. 

\section{Discussion} \label{s:discus}

\subsection{Comparison with other numerical methods}

The numerical method presented in this article is the
only method for computing relativistic binaries in
which the computational domains extends to infinity,
thereby enabling to impose exact boundary conditions on the gravitational
field equations. All the other methods \cite{BaumgCSST98,MarroMW99,UryuE00}
employ finite computational boxes. Our experience from
calculations of single rotating neutron
stars show that the finite size of the computational domain can
result in some loss of accuracy (see Ref.~\cite{NozawSGE98} for a discussion
of this point). 

Besides, thanks to the splitting of the metric quantities in a part described 
on the domains centered on star 1 and a part described on 
the domains centered on star 2, we can describe without any loss of 
accuracy very distant stars. In fact we can recover the spherical
limit when the stars have very large separations, contrary to all other 
numerical methods which are losing resolution when the stars are put
farther apart (see for instance the discussion in Sect.~V.A of
Ref.~\cite{BaumgCSST98}).

We are using surface-fitted spherical coordinates, which by construction
are well adapted to describe the stellar fluid interiors. As it can be
seen on Figs.~\ref{f:lapse}-\ref{f:extrinsic}, these coordinate systems,
which are centered on one of the two stars, are also well adapted to the
description of the metric quantities, because these latter are maximum 
at the location of the stars. Baumgarte et al. \cite{BaumgCSST98} and
Marronetti et al. \cite{MarroMW99} use instead Cartesian coordinates
in a single domain (``box''). Closer to our approach, 
Uryu \& Eriguchi \cite{UryuE00} developed a multi-domain method with
surface-fitted spherical coordinates, which enable them to treat 
precisely the fluid interiors of the stars. However, for the gravitational field
they use a single spherical coordinate system which is centered
at the system center of mass. 

As far as irrotational binaries are concerned, we payed a special attention
to the resolution of the equation for the velocity potential $\Psi$.
First we solve numerically only for a small part $\Psi_0$ of $\Psi$, thereby
reducing the numerical error. Second we let appear in the equation for $\Psi_0$ 
a partial differential operator which is invertible and give as a unique
solution that with the correct behavior at the stellar surface (velocity
field tangent to the surface in the co-orbiting frame). 
The equation for $\Psi$ is instead 
solved as a Poisson equation $\Delta\Psi={\sl source}$ 
with a boundary condition at the stellar surface
by Uryu \& Eriguchi \cite{UryuE00} and Marronetti et al. 
\cite{MarroMW99,MarroMW00}.
Note that Marronetti et al. performs only an approximate treatment of
the boundary condition, which amounts to considering that the surface of the
star is an exact sphere. This is of course not valid for close configurations. 
On the contrary, thanks to the introduction of surface-fitted coordinates, 
Uryu \& Eriguchi \cite{UryuE00} have been able to treat the boundary condition
exactly. 

\subsection{Tests passed by the code}

We have performed extensive tests of the numerical code. 
In particular, we have shown that, in the Newtonian limit, our 
numerical results coincides with the semi-analytical solutions recently
obtained by Taniguchi \& Nakamura \cite{TanigN00a,TanigN00b} for 
compressible polytropic stars. The only discrepancies appeared to
be due to missing higher order terms in Taniguchi \& Nakamura's solutions
and not to some inaccuracy of the numerical code. 

Regarding relativistic configurations, no analytical solutions was 
available to compare with. In this case, we checked only by comparing
with previous numerical solutions, namely that of Baumgarte et al. 
\cite{BaumgCSST98} for synchronized binaries and Uryu \& Eriguchi
\cite{UryuE00} for irrotational binaries. The agreement is of the
order of $1 \%$. 
For the astrophysically relevant 
case of irrotational relativistic binaries, a detailed comparison 
with Uryu \& Eriguchi code \cite{UryuE00} is underway \cite{GourgU00}. 

\subsection{Future prospects}

We are currently using the method described in this article to compute
models of close binary neutron stars with various equations of states:
polytropic EOS with various polytropic indices, dense matter EOS 
resulting from recent nuclear physics calculations. In particular,
we are studying how parameters like the frequency location of the 
innermost stable orbit (if any) depends on the equation of state, 
in order to help in the interpretation of gravitational wave signals
from coalescing neutron star binaries. The results of these studies
will be published elsewhere \cite{TanigGGB00}.

\acknowledgments

We would like to thank Luc Blanchet, Brandon Carter, Pedro Marronetti,
John Stewart and Koji Uryu for useful discussions and
Masaru Shibata for suggesting us to plot Fig.~\ref{f:N_diff}. 
The code development and the numerical computations have been performed
on SGI workstations purchased thanks to a special grant from the C.N.R.S.

\appendix

\section{Link with Teukolsky's and Shibata's formulations}
\label{s:app_link}

The first integral of motion for quasiequilibrium irrotational binaries
derived by Teukolsky \cite{Teuko98} is (cf. his Eq.~(57), re-written
within our notations\footnote{Our shift vector $\mib{B}$ is the negative
of Teukolsky's $\mib{B}$.})
\be \label{e:int_prem_Teuk}
	N \left( h^2 + \mib{D} \Psi \cdot \mib{D} \Psi \right) ^{1/2}
	 + \mib{B} \cdot  \mib{D} \Psi = C \ , 
\ee
where $C$ is some constant. 
At first glance, this might look quite different from our first integral
of motion (\ref{e:int_prem_3p1}). However, if we substitute
Eq.~(\ref{e:gamma_3p1}) for $\Gamma$ in the exponential form of our first
integral (i.e. Eq.~(\ref{e:int_prem_3p1_non_log})) we get
\be
	h N \Gamma_{\rm n} ( 1 - \mib{U}\cdot\mib{U}_0 ) = {\rm const}.
\ee
By means of Eqs.~(\ref{e:U=grad_psi}) and (\ref{e:U_0}), this equations
becomes
\be
	h N \Gamma_{\rm n} + \mib{B} \cdot  \mib{D} \Psi = {\rm const}.
\ee
Finally, if we substitute Eq.~(\ref{e:Gamma_n_cov}) for $\Gamma_{\rm n}$
is this relation, we recover Eq.~(\ref{e:int_prem_Teuk}). In particular 
this shows that the constant in the right-hand side of 
Eq.~(\ref{e:int_prem_3p1_non_log}) is nothing but the constant denoted $C$
by Teukolsky \cite{Teuko98}.

The first integral of motion for quasiequilibrium irrotational binaries
derived by Shibata \cite{Shiba98} is (cf. his Eq.~(2.18), re-written
within our notations\footnote{Our $\lambda$ and $\mib{S}$ are
denoted respectively by $u^0$ and $V^i$ by Shibata \cite{Shiba98}.})
\be \label{e:int_prem_Shibata}
	{h\over \lambda} + \mib{S}\cdot\mib{D}\Psi = {\rm const},
\ee
where $\lambda$ and $\mib{S}$ are defined by the following decomposition of
the fluid 4-velocity in a part along the Killing vector $\mib{l}$ and
a part parallel to the hypersurface $\Sigma_t$: 
\be \label{e:decomp_u_Shibata}
	\mib{u} = \lambda(\mib{l} + \mib{S}) \qquad \mbox{with}
		\qquad \mib{n}\cdot\mib{S} \ .  
\ee
Now, substituting Eq.~(\ref{e:helicoidal_n}) for $\mib{l}$ in this relation
and using Eq.~(\ref{e:def_U}), we get
\be \label{e:S_Shibata}
	\mib{S} = {1\over \lambda} \Gamma_{\rm n} \mib{U} + \mib{B} 
		= {1\over \lambda h} \mib{D} \Psi + \mib{B} \ ,  
\ee
where the second equality follows from Eq.~(\ref{e:U=grad_psi}). 
Inserting Eq.~(\ref{e:S_Shibata}) into (\ref{e:decomp_u_Shibata}) and
using the normalization relation $\mib{u}\cdot\mib{u}=-1$ results in
the following expression for $\lambda$:
\be \label{e:lambda_Shibata}
	\lambda = {1\over h N} 
	\left( h^2 + \mib{D} \Psi \cdot \mib{D} \Psi \right) ^{1/2} \ . 
\ee
Finally substituting Eq.~(\ref{e:lambda_Shibata}) for $\lambda$ and
Eq.~(\ref{e:S_Shibata}) for $\mib{S}$ into Shibata's first integral of
motion (\ref{e:int_prem_Shibata}) result in Teukolsky's form of the 
integral of motion (Eq.~(\ref{e:int_prem_Teuk}) above), which shows the
equivalence of the various formulations.

\section{Numerical method to solve the elliptic equation for the
velocity potential} \label{s:app_psi}

Equation (\ref{e:psinum2}) for the part $\Psi_0$ of the velocity potential
can be written
\be \label{e:psinum_red}
	a \ulap \Psi_0 + b^i \unab_i \Psi_0 = \sigma \ , 
\ee
with
\be
	a := \zeta H \ , 
\ee
\be
	b^i := (1-\zeta H) \unab^i H + \zeta H \unab^i \beta \ , 
\ee
and
\be
 	\sigma := (W^i - W^i_0) \unab_i H
 + \zeta H \left( W^i_0 \unab_i(H-\beta) + 
  		{W^i\over \Gamma_{\rm n}} \unab_i \Gamma_{\rm n}
                \right) \ . 
\ee
Equation (\ref{e:psinum_red}) is not merely a Poisson type equation for
$\Psi_0$ because the coefficient $a$ vanishes at the surface of the star. 
It therefore deserves a special treatment. In the works of Marronetti et
al. \cite{MarroMW99,MarroMW00} and Uryu \& Eriguchi \cite{UryuE00}, 
Eq.~(\ref{e:psinum_red})
is recast as a Poisson equation\footnote{These authors
are using $\Psi$ and not $\Psi_0$ as the unknown function, but this has
no consequence on the following discussion.} $\ulap \Psi_0 = {\sl source}$, 
dividing
both sides of Eq.~(\ref{e:psinum_red}) by $a$. In order that the right-hand
side be regular, one must then impose as a boundary condition
$b^i \unab_i \Psi_0 - \sigma = 0$ at the surface of the star. 
We choose here a different approach, considering that the operator in 
Eq.~(\ref{e:psinum_red}) is not the Laplacian but instead the operator
\be
	L\Psi_0 := \alpha(1-\xi^2) \Delta_\xi \Psi_0 
			 + \beta \xi {\partial \Psi_0 \over \partial \xi}
					\ , 
\ee
where $\alpha$ and $\beta$ are two constants,
$\xi \in[0,1]$ is the computational radial coordinate introduced in the
mapping (\ref{e:map_r_xi_1}), and $\Delta_\xi$ is an operator which, expressed
in terms of the computational coordinates $(\xi,\theta',\varphi')$ has
the same structure than the Laplacian operator:
\be
   \Delta_\xi \Psi_0 := {1\over\xi^2} {\partial \over \partial \xi} \left(
		\xi^2 {\partial \Psi_0\over \partial \xi} \right)
			+ {1\over\xi^2\sin\theta'} 
		{\partial \over \partial \theta'} \left(
		\sin\theta' {\partial \Psi_0\over \partial \theta'} \right)
			+ {1\over\xi^2\sin^2\theta'}  
		{\partial^2 \Psi_0\over \partial {\varphi'}^2} \ . 
\ee
Here we assume that there is only one domain covering the star, i.e.
that $M_{\langle 1 \rangle} = M_{\langle 2 \rangle} = 1$, so that the
surface of the star is given by $\xi=1$. Equation~(\ref{e:psinum_red})
is then re-written as
\be \label{e:psinum_final}
  L\Psi_0 = \sigma + \alpha(1-\xi^2) \Delta_\xi \Psi_0 - a \ulap \Psi_0
   + \beta \xi {\partial \Psi_0 \over \partial \xi} - b^i \unab_i \Psi_0 \ . 
\ee
The basic idea is to solve this equation by iterations, considering 
the whole right-hand as a source term and using 
the previous step value of $\Psi_0$ in it. One must also choose the
constants $\alpha$ and $\beta$ so that the term
$\alpha(1-\xi^2) \Delta_\xi \Psi_0$
[resp. $\beta \xi {\partial \Psi_0 /\partial \xi}$] is as close as possible
to $a \ulap \Psi_0$
[resp. $b^i \unab_i \Psi_0$]. We opt for the following choices:
\be
	\alpha = \left. a \left( {\partial r \over \partial \xi} \right) ^2  
			\right| _{\xi=0}
\ee
and 
\be
	\beta = - \max \left| b^r 
			\left( {\partial r \over \partial \xi} \right) ^2
					 \right| \ ,
\ee
where $b^r$ is the $r$-component of the vector $b^i$. 
In solving Eq.~(\ref{e:psinum_final}) by iterations, we introduce the following
relaxation
\be
  \Psi_0^J \leftarrow  \lambda  \Psi_0^J + (1-\lambda) \Psi_0^{J-1}  \ , 
\ee
where  $J$ (resp. $J-1$) labels the current step (resp. previous step)
and $\lambda$ is the relaxation factor, typically chosen to be $0.5$. 

At each step of the iteration, Eq.~(\ref{e:psinum_final}) is solved by
the following spectral method. First an expansion in spherical harmonics
$Y_\ell^m(\theta',\varphi')$ is performed, so that Eq.~(\ref{e:psinum_final})
becomes equivalent to a set of ordinary differential equations [one equation
for each couple $(\ell,m)$]:
\be
	L_\ell f(\xi)  =  s(\xi)             \label{e:L_lm} 
\ee
where $f(\xi)$ and $s(\xi)$ are the $(\ell,m)$ coefficient
of $\Psi_0$ and of the whole right-hand side of (\ref{e:psinum_final}) 
respectively and $L_\ell$ is the following differential operator:
\be
	L_\ell f  :=  
	\alpha(1-\xi^2) \left[ {d^2 f \over d\xi^2}
	+ {2\over \xi} {df\over d\xi} - {\ell(\ell+1)\over \xi^2} f \right]
	+ \beta \xi {df\over d\xi}  \ .	
\ee
 
Since the source $s(\xi)$ vanishes for $\ell=0$, we treat only the
case $\ell > 0$. 
Regularity properties at the origin ($\xi=0$) imply that $f(\xi)$ and
$s(\xi)$ should be expandable in even (resp. odd) Chebyshev polynomials
$T_n(\xi)$ for $\ell$ even (resp. odd).
Due to the division by $\xi$ and $\xi^2$, the differential operator $L_\ell$ is
singular on Chebyshev polynomials at $\xi=0$, except for $\ell=1$. 
Therefore, instead of Chebyshev polynomials, we use the 
following polynomials
$P_n(\xi)$ as a expansion basis for $f$ ($N$ is the total number of 
coefficients in the Chebyshev expansions, denoted $N_r^{\langle a\rangle}(0)$
in Sect.~\ref{s:spectr})
\begin{itemize}
\item for $\ell$ even : 
$P_n(\xi) := T_{2n}(\xi) + T_{2n+2}(\xi) = 2\xi \, T_{2n+1}(\xi)$,\quad
 $0\leq n\leq N-2$; 
\item for $\ell=1$ : $P_n(\xi) := T_{2n+1}(\xi)$,\quad  $0\leq n\leq N-1$;
\item for $\ell$ odd and $\ell>1$: 
  $P_n(\xi) := (2n+3) T_{2n+1}(\xi) + (2n+1) T_{2n+3}(\xi)$,\quad
   $0\leq n \leq N-2$.
\end{itemize} 
The operator on the left-hand side of Eq.~(\ref{e:L_lm}) is regular for
each of the polynomials $P_n(\xi)$ (such a basis
is called a {\em Galerkin basis}). 

We thus consider the differential operator $L_\ell$ acting
\begin{itemize}
\item for $\ell$ even : from the $N-1$ dimensional vectorial space span by the
polynomials $P_n(\xi)$ ($0\leq n\leq N-2$) to the $N-1$ dimensional
vectorial space span by the polynomials $T_{2n}(\xi)$ ($0\leq n\leq N-2$); 
\item for $\ell=1$ : from the $N$ dimensional vectorial space span by the
polynomials $P_n(\xi)= T_{2n+1}(\xi)$ ($0\leq n\leq N-1$) to itself;
\item for $\ell$ odd and $\ell>1$: from the $N-1$ dimensional vectorial 
space span by the polynomials $P_n(\xi)$ ($0\leq n\leq N-2$) to the 
$N-1$ dimensional vectorial space span by the polynomials 
$T_{2n+1}(\xi)$ ($0\leq n\leq N-2$). 
\end{itemize} 
The operator $L_\ell$ is then one-to-one (isomorphism) between these
vectorial spaces. This means that the only homogeneous solution is 
zero. Otherwise stated, for each $\ell$ there is a unique solution to 
Eq.~(\ref{e:L_lm}) in the vectorial space spans by the $P_n(\xi)$'s.
To find this solution, we transform the matrix $A_{ij}$ of $L_\ell$ in the
above bases into a banded matrix
by means of the following linear combinations:
\begin{itemize}
\item for $\ell$ even : 
\begin{eqnarray}
  \bar A_{ij} & = & {1\over i+1} [ A_{ij} - A_{(i+1)j} ]
 \quad \mbox{for} \ 0 \leq i \leq N-3 \ ; \\
  \tilde A_{ij} & = & \bar A_{ij} - \bar A_{(i+2)j}  
 \quad \mbox{for} \ 0 \leq i \leq N-5 \ . 
\end{eqnarray}
\item for $\ell$ odd : 
\begin{eqnarray}
  \bar A_{ij} & = & {1\over i+1} [ (1+\delta_0^i) A_{ij} - A_{(i+2)j} ]
 \quad \mbox{for} \ 0 \leq i \leq N-3 \ ; \\
  \tilde A_{ij} & = & \bar A_{ij} - \bar A_{(i+2)j}  
 \quad \mbox{for} \ 0 \leq i \leq N-5 \ . 
\end{eqnarray}
\end{itemize}
Since the resulting matrix $\tilde A_{ij}$ has at most 5 bands, 
the linear system is easily and CPU-efficiently solved to get the coefficients 
of the solution $f$ in the
basis of the polynomials $P_n(\xi)$. A simple combination is then performed 
on these coefficients to get the coefficients on the usual Chebyshev
bases. 

A very discriminating test of this numerical technique, namely the
evaluation of the tiny $z$-component of the velocity
field resulting from $\Psi$, is presented
in Figs.~\ref{f:velo1}-\ref{f:velo3} (Sect.~\ref{s:comp_anal}).

\end{document}